# Approach-Level Real-Time Crash Risk Analysis for Signalized Intersections

Jinghui Yuan*, Mohamed Abdel-Aty

Department of Civil, Environmental & Construction Engineering, University of Central Florida, Orlando, FL 32816, USA

* Corresponding author. Tel: +1-407-881-4706. E-mail address: jinghuiyuan@knights.ucf.edu

**ABSTRACT:**

Intersections are among the most dangerous roadway facilities due to the complex traffic conflicting movements and frequent stop-and-go traffic. However, previous intersection safety analyses were conducted based on static and highly aggregated data (e.g., annual average daily traffic (AADT), annual crash frequency). These aggregated data may result in unreliable findings simply because they are averages and cannot represent the real conditions at the time of crash occurrence. This study attempts to investigate the relationship between crash occurrence at signalized intersections and real-time traffic, signal timing, and weather characteristics based on 23 signalized intersections in Central Florida. The intersection and intersection-related crashes were collected and then divided into two types, i.e., within intersection crashes and intersection entrance crashes. Bayesian conditional logistic models were developed for these two kinds of crashes, respectively. For the within intersection models, the model results showed that the through volume from "A" approach (the traveling approach of at-fault vehicle), the left turn volume from "B" approach (near-side crossing approach), and the overall average flow ratio (OAFR) from "D" approach (far-side crossing approach), were found to have significant positive effects on the odds of crash occurrence. Moreover, the increased adaptability for the left turn signal timing of "B" approach and more priority for "A" approach could significantly decrease the odds of crash occurrence. For the intersection entrance models, average speed was found to have significant negative effect on the odds of crash occurrence. The longer average green time and longer average waiting time for the left turn phase, higher green ratio for the through phase, and higher adaptability for the through phase can significantly improve the safety performance of intersection entrance area. In addition, the average queue length on the through lanes was found to have positive effect on the odds of crash occurrence. These results are important in real-time safety applications at signalized intersections in the context of proactive traffic management.

# 1. Introduction

Intersections are among the most dangerous roadway facilities due to the complex traffic conflicting movements and frequent stop-and-go traffic. Take Florida as an example, nearly 26% of crashes happen at or influenced by intersections (including signalized and non-signalized) in 2014. Moreover, signalized intersections are generally large intersections with higher traffic volume, therefore, the safety status of signalized intersection would be even more complicated. Safety analysis for signalized intersection has been a critical research topic during past decades. Substantial efforts have been made by previous researchers to reveal the relationship between crash frequency of signalized intersections and all the possible contributing factors such as roadway geometric, signal control, and traffic characteristics, etc. (Chin and Quddus 2003, Abdel-Aty and Wang 2006, Wang *et al.* 2006, Wang *et al.* 2009, Guo *et al.* 2010, Lee *et al.* 2017, Wang and Yuan 2017, Cai *et al.* 2018a, Cai *et al.* 2018b, Wang *et al.* 2018).

More specifically, nearly all the traffic volume related variables were found to have significant positive effects on the crash frequency at signalized intersections, including total entering ADT (Poch and Mannering 1996, Chin and Quddus 2003, Abdel-Aty and Wang 2006, Guo *et al.* 2010), right-turn ADT (Poch and Mannering 1996, Chin and Quddus 2003), left-turn ADT (Poch and Mannering 1996), total ADT on major road (Wang *et al.* 2009, Dong *et al.* 2014), total ADT on minor road (Wang *et al.* 2009, Dong *et al.* 2014), left-turn ADT on major road (Guo *et al.* 2010), through ADT on minor road (Guo *et al.* 2010). However, Guo *et al.* (2010) found that the through ADT on major road and the left-turn ADT on minor road are significantly negatively associated with the crash frequency at signalized intersections. Moreover, Wang *et al.* (2009) investigated the relationship between LOS and safety at signalized intersections. They found that LOS D is a desirable level which is associated with less total crashes, rear-end and sideswipe crashes, as well as right-angle and left-turn crashes. Xie *et al.* (2013) investigated the safety effect of corridor-level travel speed, they found that the high speed corridor may results in more crashes at the signalized intersections. Similarly, the speed limit of the corridor was found to be significantly positively correlated with the crash frequency of the signalized intersections (Poch and Mannering 1996, Abdel-Aty and Wang 2006, Wang *et al.* 2009, Guo *et al.* 2010, Dong *et al.* 2014).

With respect to the geometric design, number of lanes, median width, and intersection sight distance et al. were found to have significant effects on the crash frequency of signalized intersections. More specifically, the number of lanes was found to be positively correlated with the crash frequency of signalized intersections (Poch and Mannering 1996, Abdel-Aty and Wang 2006, Guo *et al.* 2010, Dong *et al.* 2014). Median width and intersection sight distance was also found to have positive effect on the crash frequency(Chin and Quddus 2003). Moreover, Abdel-Aty and Wang (2006) found that the existence of exclusive right-turn lanes could significantly decrease the crash frequency.

In terms of signal control characteristics, the adaptive signal control was found to have significant lower crash frequency than the pre-timed signal control (Chin and Quddus 2003). The number of phase was found to be positively associated with the crash frequency of signalized intersections (Poch and Mannering 1996, Chin and Quddus 2003, Xie *et al.* 2013). The left-turn protection could significantly improve the safety performance of the signalized interaction (Poch and Mannering 1996, Chin and Quddus 2003, Abdel-Aty and Wang 2006). However, Abdel-Aty and Wang (2006) found that the left-turn protection on minor roadway tends to increase the crash frequency of signalized intersection. Surprisingly, Guo *et al.* (2010) found that the coordinated intersections are more unsafe than the isolated ones. They explained it as the travel speed is higher for coordinated intersections because of the green wave, which may results in more crashes.

However, these studies were conducted based on static and highly aggregated data (e.g., Annual Average Daily Traffic (AADT), annual crash frequency). These aggregated data limit the reliability of the findings simply because they are averages and cannot reflect the real conditions at the time of crash occurrence. With the rapid development of traffic surveillance system and detection technologies, real-time traffic data are not only available on freeways and expressways but also on urban arterials (including road segments and intersections). During the past decade, an increasing number of studies have investigated the crash likelihood on freeways by using real-time traffic and weather data (Oh *et al.* 2001, Lee *et al.* 2003, Abdel-Aty *et al.* 2004, Zheng *et al.* 2010, Abdel-Aty *et al.* 2012, Ahmed *et al.* 2012a, Xu *et al.* 2013a, Xu *et al.* 2013b, Yu and Abdel-Aty 2014, Yu *et al.* 2014, Basso *et al.* 2018, Theofilatos *et al.* 2018). It is worth noting that Theofilatos *et al.* (2018) investigated crash occurrence by utilizing real-time traffic data while considering that the number of crashes is very few, and they could be considered as rare events. In this context, they compared the model results of different crash to non-crash ratio (1:10 and full sample of non-crash events) by using two different statistical models (bias correction and firth model), respectively. It was found that the two methods have different advantages and disadvantages, and the choice of the most appropriate method depends on several criteria. Also, Basso *et al.* (2018) developed real-time crash prediction model for urban expressway based on the original unbalanced data, rather than artificially balanced data by using Synthetic Minority Over-sampling Technique (SMOTE). They claimed that their model performance are among the best in the literature.

However, little research has been conducted on the real-time safety of urban arterials (Theofilatos 2017, Theofilatos *et al.* 2017, Yuan *et al.* 2018), especially signalized intersections (Mussone *et al.* 2017). Mussone *et al.* (2017) examined the factors which may affect the crash severity level at intersection based on real-time traffic flow and environmental characteristics, and they found that the real-time traffic flow characteristics have a relevant role in predicting crash severity. However, they didn't consider the crash likelihood at intersections, which means that the effects of real-time traffic flow and environmental characteristics on the crash likelihood at intersections are still unclear.

Moreover, the conflicting traffic movements at signalized intersection are temporally separated by traffic signals. Therefore, signal timing plays a very important role in the intersection safety, especially when the adaptive signal control technology was widely adopted on major urban arterials. Adaptive signal control technology optimize signal timing plans in real-time, it was found to have significant effects in reducing stops and delays (Khattak *et al.* 2018a) and improving traffic safety (Chin and Quddus 2003, Khattak *et al.* 2018b). However, the safety effect of real-time signal status has never been considered, while improper signal timing may result in dangerous situation. Therefore, the relationship between real-time signal timing and intersection safety need to be further investigated.

On the other hand, with the rapid development of connected vehicle technologies in recent years, it is feasible for us to implement efficient proactive traffic management strategies at intersections, e.g., dynamic message sign (DMS) to show the real-time crash risk for the downstream intersections, and vehicle-level optimal speed advisory through vehicle-to-infrastructure (V2I) communication (Yue *et al.* 2018). In this context, an efficient and reliable real-time crash risk predictive algorithm for intersections is required. However, traditional intersection safety analysis were usually conducted by modeling historical crash frequency with geometric, AADT, and static signal control characteristics, which ignore the impacts of real-time traffic environment (e.g., traffic and weather) when crashes occur.

To the best of the authors' knowledge, there have been no studies done on the real-time crash risk at signalized intersections. To bridge this gap, this study aims to investigate the relationship between crash likelihood at signalized intersections and real-time traffic, signal timing, and weather characteristics by utilizing data from multiple sources, i.e., Bluetooth, weather, and adaptive signal control datasets.

## 2. Data Preparation

There are 23 intersections chosen from four urban arterials in Orlando, Florida, as shown in Figure 1. A total of four datasets were used: (1) crash data from March, 2017 to March, 2018 provided by Signal Four Analytics (S4A); (2) travel speed data collected by 23 IterisVelocity Bluetooth detectors installed at 23 intersections; (3) signal phasing and 15-minute interval traffic volume provided by 23 adaptive signal controllers; (4) weather characteristics collected by the nearest airport weather station.

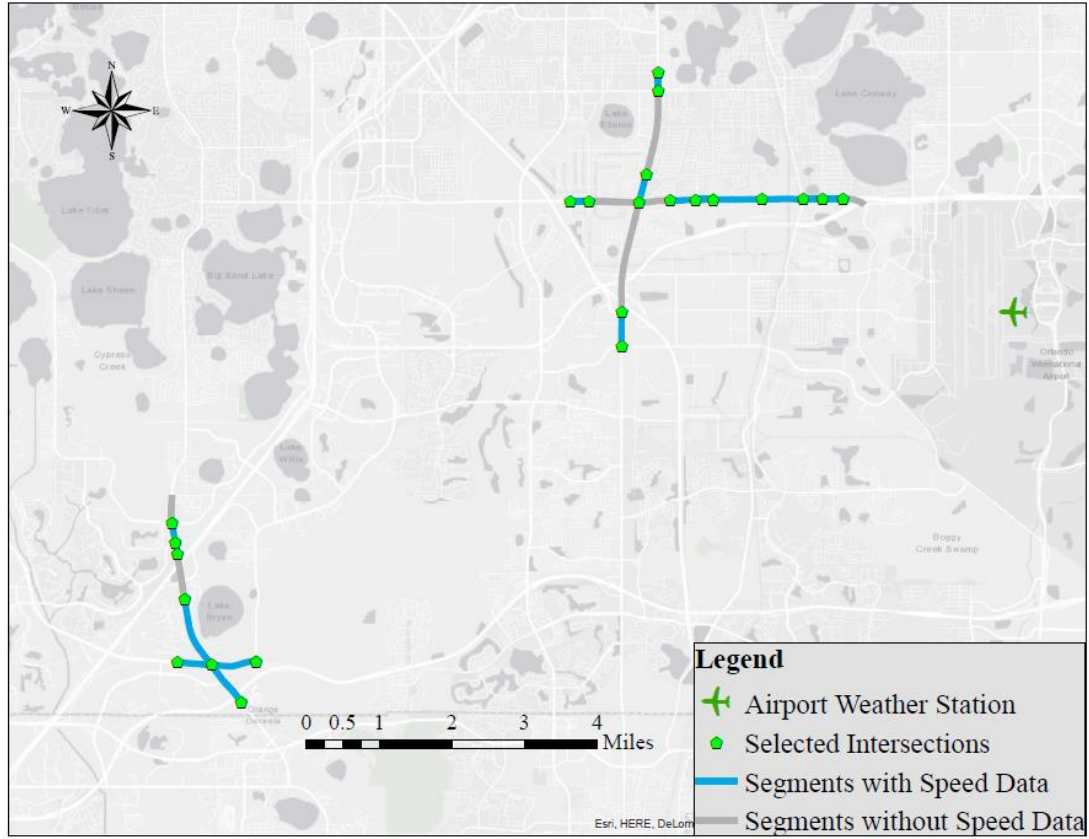

Figure 1. Layout of Selected Intersections

S4A provides detailed crash information, including crash time, coordinates, severity, type, weather condition, etc. In terms of the crash time information, there are three kinds of time information for each crash, i.e. time of crash occurrence, time reported, and time dispatched. Only the time of crash occurrence was utilized in this study, and the difference between this recorded crash time and the actual crash time is supposed to be within 5 minutes since there exist several efficient and accurate technologies for the police officer to identify the accurate time of crash occurrence, e.g. closed-circuit television cameras and mobile phones.

First, all crashes occurred at intersection or influenced by intersection (within 250 feet of intersection) from March, 2017 to March, 2018 were collected. Second, all the single-vehicle crashes and the crashes under the influence of alcohol and drugs were excluded, since these kinds of crashes are usually not attributed to the real-time traffic and signal characteristics which are the focus of this study. After that, a total of 803 crashes remained and these crashes were divided into three types based on their location, which are within intersection area, intersection entrance area, and intersection exit area, as shown in Figure 2. There are 446 (55.54%) crashes that had occurred within intersection, 264 (32.88%) crashes that had occurred in the intersection entrance area, and 93 (11.58%) crashes that had occurred in the intersection exit area. In terms of the sample size, only within intersection crashes and intersection entrance crashes were utilized in this study.

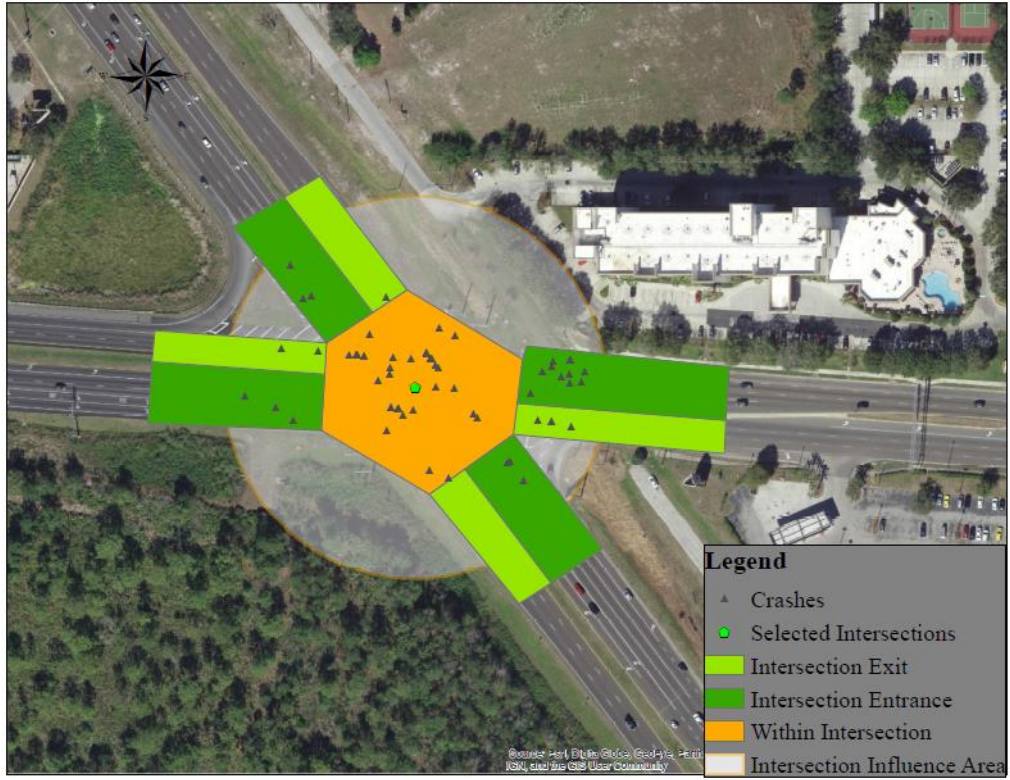

Figure 2. Illustration of Three Types of Intersection Crash Location

Before collecting the real-time traffic and signal timing variables for each crash, two preprocess steps were conducted: First, identify the travel direction of the at-fault vehicle in each crash based on the attribute of "Crash Type Direction", and then rename the approach of at-fault vehicle as "A" approach; Second, retrieve the travel direction of the other three approaches based on the nomenclature in Figure 3, and then rename them as "B", "C", and "D" approaches, respectively. After this preprocessing, all the relationship between crash location and intersection approaches were consistent, i.e., the travel approach of the at-fault vehicles for all crashes were named as "A" approach and all the other corresponding approaches were named as "B", "C", and "D" approaches according to the nomenclature. For the within intersection crash and non-crash events, the real-time traffic and signal timing data were collected from four approaches, while for the intersection entrance crash and non-crash events, only the data from "A" approach were collected.

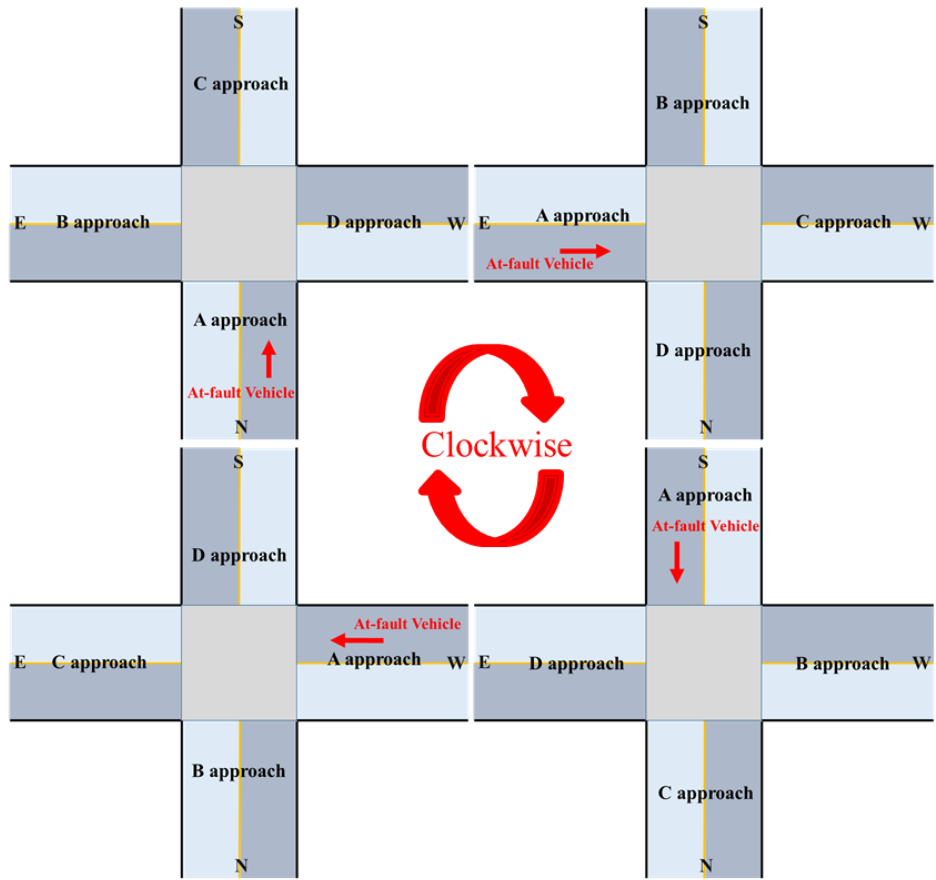

Figure 3. The Nomenclature of the Four Approach ("A", "B", "C", and "D")

Matched case-control design was employed in this study to explore the effects of traffic, signal, and weather related variables while eliminating the effects of other confounding factors through the design of study. For each crash, four confounding factors, i.e., intersection ID, crash location type (within intersection or intersection entrance), time of day, and day of week, were selected as matching factors. Therefore, a group of non-crash events could be identified by using these matching factors and then a specific number of non-crash events could be randomly selected from this group of non-crash events for every crash event. The number of non-crash events *m* corresponding to a crash event is preferred to be fixed in the entire analysis. As stated in Hosmer Jr *et al.* (2013), the value of *m* was commonly chosen from one to five. Moreover, Abdel-Aty *et al.* (2004) found that there is no significant difference when *m* changing from one to five. Therefore, the control-to-case ratio of 4:1 was adopted in this study, which is consistent with previous studies (Abdel-Aty *et al.* 2008, Zheng *et al.* 2010, Ahmed *et al.* 2012b, Ahmed and Abdel-Aty 2012, Xu *et al.* 2012, Ahmed and Abdel-Aty 2013, Shi and Abdel-Aty 2015, Yu *et al.* 2016). Consequently, 4 non-crash events from the same intersection, crash location type, time of day, and day of week were randomly selected for each crash event. Figure 4 shows an example of the matched case control design for the within intersection crash event. Besides, the non-crash events were selected only when there is no crashes occurring within 3 hours before or after the non-crash event on the same location.

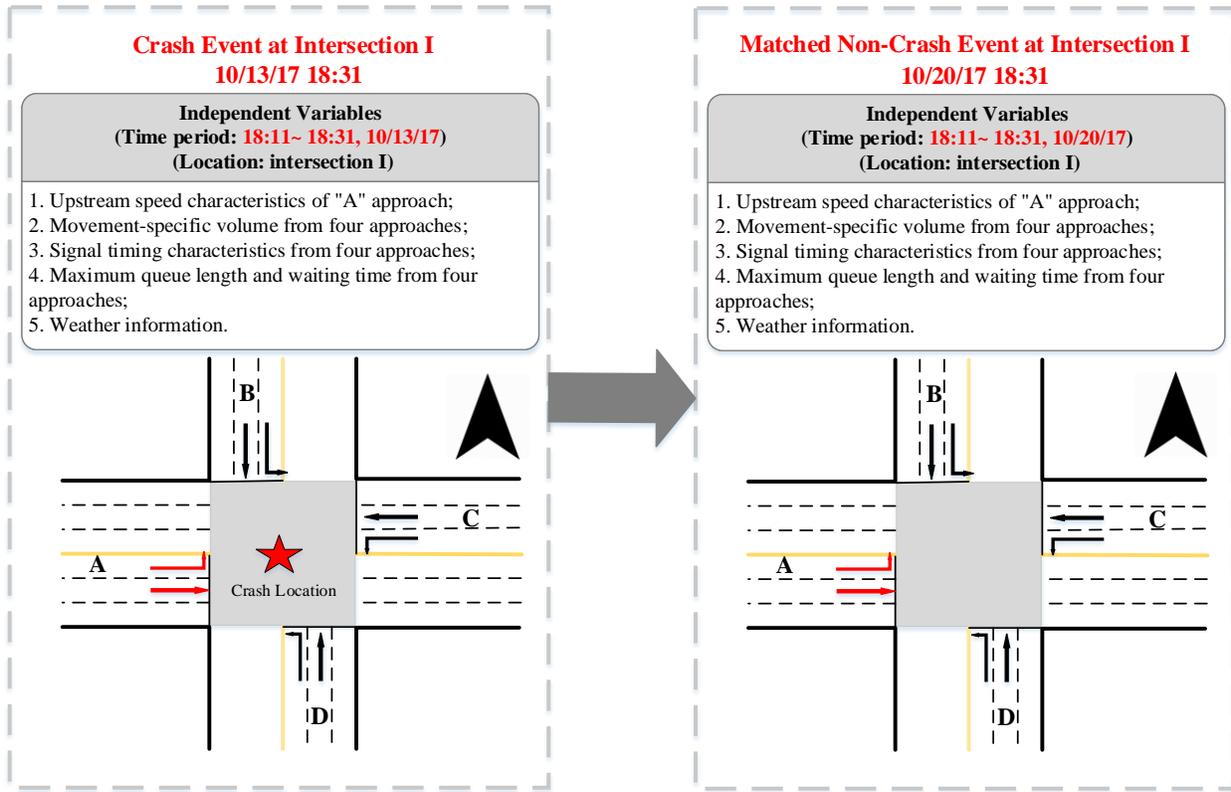

Figure 4. Illustration of Matched Case-Control Design for the Within-Intersection Crashes

The real-time traffic and signal timing data for both crash and non-crash events were extracted for a period of 20 minutes (divided into four 5-minute time slices) prior to crash occurrence. For example, if a crash event $i$ occurred within intersection at 18:31, the corresponding traffic and signal timing data from 18:11 to 18:31 were extracted and named as time slice 4, 3, 2, and 1, respectively. As shown in Figure 5, the traffic and signal timing data collection for different crash location are different. For the within-intersection crashes, all the traffic and signal timing variables from four approaches were collected. However, for the intersection entrance crashes, data were collected only from the "A" approach.

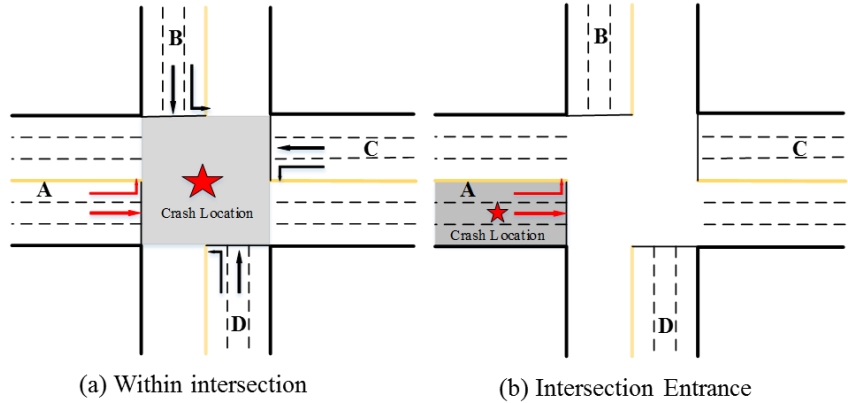

(a) Within intersection     (b) Intersection Entrance

Figure 5. Schematic Figure of Crash Location and Data Collection

Speed data were provided by the 23 IterisVelocity Bluetooth detectors, which measure the space-mean speed of a specific segment, as shown in Figure 6. Bluetooth detectors can only detect the vehicles equipped with Bluetooth device which is working at discoverable mode. The space-mean speed of each vehicle on a specific segment is calculated as the segment length divided by the travel time of each detected vehicle on the segment based on the detection data of two Bluetooth detectors located at the two contiguous intersections. In this study, speed data, including average speed and speed standard deviation, were only collected for the segment of "A" approach, which represents the traveling segment of the at-fault vehicle. Moreover, since all the Bluetooth detectors are installed on the major arterials, therefore, only the major approaches were provided with the real-time traffic speed data. In this context, all the intersection entrance crashes included in the final datasets were occurred on the major approach, and all the at-fault vehicles of the within intersection crashes were coming from the major approach.

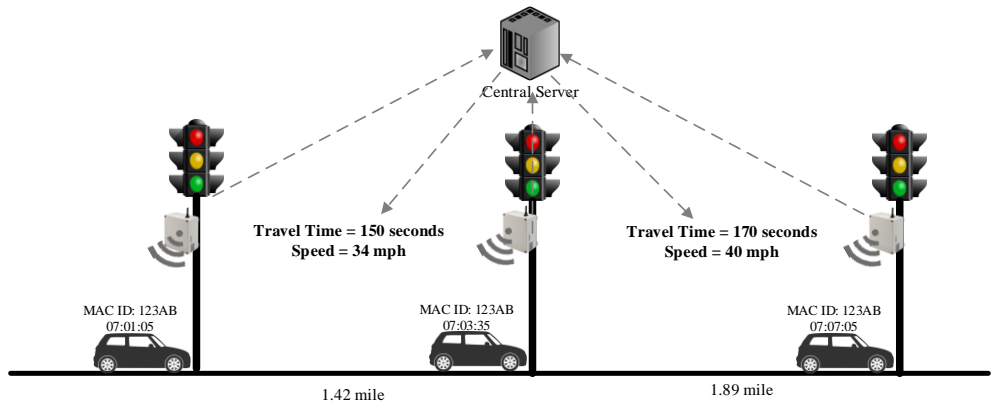

Figure 6. Illustration of Bluetooth Data Collection (Yuan *et al.* 2018)

Adaptive signal controllers archive the real-time signal timing and lane-specific 15-minute aggregate traffic volume data. The lane-specific 15-minute aggregated traffic volume data are collected by the video detectors, which are installed for the adaptive signal controller to detect the real-time volume, queue length and waiting time. Since the right-turn vehicles are unprotected at the intersection, the traffic volume data only include the through and left-turn vehicles. The lane-specific traffic volume for each time slice (5-minute) was calculated based on the assumption that the traffic volume within 15-minute interval are evenly distributed. Moreover, the variation in traffic flow across lanes in the form of overall average flow ratio (OAFR) were considered in this study. The OAFR was proposed by Lee *et al.* (2006) to represent a surrogate measure of the lane change frequency within all lanes. The OAFR is calculated as the geometric mean of the modified average flow ratio (AFR) of all lanes, while the modified AFR is calculated as the ratio of the average flow in the adjacent lanes ($i-1, i+1$) to the average flow in the subject lane ($i$), as shown in Eq.1.

$$AFR_i(t) = \frac{V_{i-1}(t)}{V_i(t)} \times \frac{NL_{i-1,i}(t)}{NL_{i-1,i}(t) + NL_{i-1,i-2}(t)} + \frac{V_{i+1}(t)}{V_i(t)} \times \frac{NL_{i+1,i}(t)}{NL_{i+1,i}(t) + NL_{i+1,i+2}(t)} \quad (1)$$

Where $V_i(t)$ is average flow in the subject lane $i$ during time interval $t$; $V_{i-1}(t)$ and $V_{i+1}(t)$ are the average flow in the adjacent lanes $i-1$ and $i+1$, respectively during time interval $t$; $NL_{i-1,i}(t)$ is the number of lane changes from lane $i-1$ to lane $i$, if lane $i-1$ exists, during time interval $t$; similarly, $NL_{i-1,i-2}(t)$, $NL_{i+1,i}(t)$, and $NL_{i+1,i+2}(t)$ represent the number of lane changes from lane $i-1$ to $i-2$, $i+1$ to $i$, and $i+1$ to $i+2$ during time interval $t$, respectively. Because the fractions of the number of lane change from lane $i-1$ to lane $i$ and $i-2$, as well as the fractions from lane $i+1$ to lane $i$ and $i+2$, were unknown in this study, they were assumed to be equal, which is in line with Lee *et al.* (2006).

It is worth noting that the OAFR calculated by Lee *et al.* (2006) as the geometric mean of the modified average flow ratio (AFR) of all lanes is only appropriate for the segment with lane number greater than 3. If the total lane number is 2, the calculated OAFR will always be 0.5 ($\sqrt[2]{\frac{V_1(t)}{V_2(t)} \times 0.5 \times \frac{V_2(t)}{V_1(t)} \times 0.5}$), no matter with the real flow variation between these two lanes. Therefore, the OAFR in this study was calculated as the arithmetic mean of the modified AFR ($\frac{1}{n}\sum_{i=1}^{n} AFR_i(t)$).

Three weather related variables (weather type, visibility, and hourly precipitation) were collected from the nearest airport weather station, which is located at the Orlando international airport (as shown in Figure 1). Since the weather data is not recorded continuously, once the weather condition changes and reaches a preset threshold, a new record will be added to the archived data. Therefore, for each specific crash, based on the reported crash time, the closest weather record prior to the crash time has been extracted and used as the crash time weather condition, which is identical for four time slices (Chung *et al.* 2018). A cross table was made to validate the weather type information extracted from weather station and the weather condition recorded in the crash report, results indicated that the consistency ((True positive + True negative)/Total sample size) between weather station and crash report is around 92%. Therefore, all the weather information for both crash and non-crash events were extracted from the airport weather station data.

After the above data collection process, the final dataset for the within intersection area includes 470 observations (94 crash events and 376 non-crash events), while the final dataset for the intersection entrance area includes 425 observations (85 crash events and 340 non-crash events). The summary statistics of within intersection and intersection entrance datasets are as shown in Table 1 and Table 2, respectively.

Table 1. Summary of Variables Descriptive Statistics for the Within Intersection Area (Crash and Non-crash Events)

| Variable | Time Slice | Description | Crash Events | | Non-Crash Events | |
|---|---|---|---|---|---|---|
| | | | Mean (Std) | (Min, Max) | Mean (Std) | (Min, Max) |
| **Avg_speed** | 1 | Average speed on the upstream segment of "A" approach within 5-minute interval (mph). | 25.69 (9.52) | (5.00, 45.57) | 26.94 (10.42) | (4.75, 54.00) |
| | 2 | | 27.8 (10.32) | (6.20, 51.67) | 27.11 (10.29) | (6.50, 56.00) |
| | 3 | | 27.43 (10.32) | (5.00, 52.00) | 27.04 (10.37) | (6.42, 53.00) |
| | 4 | | 26.9 (10.33) | (5.50, 54.00) | 27.10 (10.27) | (4.60, 54.75) |
| **Std_speed** | 1 | Speed standard deviation on the upstream segment of "A" approach within 5-minute interval (mph). | 10.59 (4.70) | (0.00, 20.92) | 9.83 (5.15) | (0.00, 27.58) |
| | 2 | | 9.39 (4.69) | (0.71, 21.21) | 10.15 (5.34) | (0.00, 36.77) |
| | 3 | | 10.05 (5.09) | (0.00, 23.33) | 10.14 (5.49) | (0.00, 36.06) |
| | 4 | | 10.62 (5.26) | (0.00, 22.19) | 10.12 (5.34) | (0.00, 26.87) |
| **A_Vol_LT** | 1 | Left turn volume of "A" approach within 5-minute interval (vehicle). | 24.84 (24.14) | (0.00, 133.67) | 22.26 (22.48) | (0.00, 186.00) |
| | 2 | | 24.44 (21.93) | (0.00, 125.67) | 21.99 (20.96) | (0.00, 186.00) |
| | 3 | | 23.94 (20.84) | (0.00, 125.67) | 21.77 (19.67) | (0.00, 177.67) |
| | 4 | | 24.50 (25.00) | (0.00, 192.00) | 22.24 (21.48) | (0.00, 177.67) |
| **A_Vol_Th** | 1 | Through volume of "A" approach within 5-minute interval (vehicle). | 112.30 (50.86) | (0.00, 298.33) | 106.09 (54.43) | (0.00, 481.33) |
| | 2 | | 113.73 (49.08) | (0.00, 298.33) | 106.24 (51.07) | (0.00, 404.00) |
| | 3 | | 109.82 (48.26) | (0.00, 259.33) | 104.89 (50.61) | (0.00, 369.80) |
| | 4 | | 113.65 (63.11) | (0.00, 416.00) | 105.79 (54.53) | (0.00, 405.33) |
| **A_OAFR** | 1 | Overall average flow ratio of "A" approach within 5-minute interval. | 1.33 (1.22) | (0.94, 11.29) | 1.40 (2.25) | (0.94, 38.88) |
| | 2 | | 1.48 (1.79) | (0.95, 11.29) | 1.56 (3.09) | (0.94, 37.27) |
| | 3 | | 1.69 (3.26) | (0.94, 29.42) | 1.58 (2.7) | (0.94, 30.28) |
| | 4 | | 1.75 (3.38) | (0.94, 29.42) | 1.54 (2.64) | (0.94, 30.28) |
| **A_LT_GreenRatio** | 1 | Ratio of left turn green time on "A" approach within 5-minute interval (%). | 13.94 (7.27) | (2.33, 34.67) | 13.54 (7.82) | (1.67, 44.67) |
| | 2 | | 14.16 (6.63) | (2.67, 31.33) | 14.39 (8.17) | (1.67, 41.33) |
| | 3 | | 14.15 (6.59) | (3.00, 35.67) | 13.79 (7.17) | (1.67, 36.33) |
| | 4 | | 13.55 (7.74) | (2.33, 36.00) | 13.99 (7.81) | (0.67, 39.67) |
| **A_LT_Avg_Green** | 1 | Average length of left turn green phase on "A" approach within 5-minute interval (second). | 18.41 (9.49) | (4.00, 46.00) | 18.45 (9.74) | (2.50, 50.00) |
| | 2 | | 18.82 (9.16) | (6.40, 41.00) | 18.86 (10.67) | (2.00, 67.00) |
| | 3 | | 18.26 (10.05) | (4.50, 57.00) | 18.34 (9.61) | (4.00, 61.00) |
| | 4 | | 17.74 (9.18) | (3.50, 47.00) | 18.55 (9.66) | (2.00, 64.00) |
| **A_LT_Std_Green** | 1 | Standard deviation of the length of left turn green phase on "A" approach within 5-minute interval (second). | 5.90 (6.34) | (0.00, 40.20) | 5.90 (5.9) | (0.00, 31.11) |
| | 2 | | 5.91 (5.64) | (0.00, 36.77) | 5.13 (5.24) | (0.00, 34.65) |
| | 3 | | 6.52 (6.14) | (0.00, 26.87) | 6.69 (6.52) | (0.00, 43.84) |
| | 4 | | 5.28 (4.71) | (0.00, 21.21) | 6.32 (5.62) | (0.00, 31.11) |
| **A_LT_Avg_Queue** | 1 | Average left turn queue length at the beginning of left turn green phase on "A" approach (vehicle). | 8.39 (6.54) | (1.00, 33.33) | 8.90 (7.40) | (0.00, 47.00) |
| | 2 | | 8.56 (6.07) | (0.75, 33.33) | 8.74 (7.03) | (0.00, 46.00) |
| | 3 | | 9.58 (7.84) | (0.33, 40.00) | 8.59 (6.70) | (0.00, 45.00) |
| | 4 | | 9.03 (7.34) | (0.00, 40.00) | 9.10 (7.46) | (0.00, 45.00) |
| **A_LT_Avg_Wait** | 1 | Average left turn maximum waiting time at the beginning of | 94.69 (45.35) | (0.50, 167.50) | 97.25 (48.07) | (0.00, 266.00) |
| | 2 | | 95.16 (45.79) | (0.50, 179.00) | 97.96 (49.07) | (0.00, 241.00) |
| | 3 | | 96.72 (51.78) | (0.40, 279.00) | 97.71 (49.14) | (0.00, 246.5) |
| | 4 | left turn green phase on "A" approach (vehicle). | 98.59 (48.58) | (2.50, 169.50) | 95.96 (49.48) | (0.00, 284.00) |
| **A_TH_GreenRatio** | 1 | Ratio of through green time on "A" approach within 5-minute interval (%). | 44.52 (16.00) | (13.67, 85.67) | 44.3 (16.25) | (7.33, 88.33) |
| | 2 | | 44.30 (15.40) | (14.67, 85.00) | 43.91 (15.56) | (6.33, 91.67) |
| | 3 | | 43.98 (16.00) | (15.33, 84.67) | 43.3 (16.17) | (12.00, 83.67) |
| | 4 | | 43.39 (16.37) | (10.67, 89.67) | 43.48 (16.51) | (8.33, 88.67) |
| **A_TH_Avg_Green** | 1 | Average length of through green phase on "A" approach within 5-minute interval (second). | 28.88 (17.99) | (11.2, 105.00) | 29.42 (19.58) | (9.64, 128.00) |
| | 2 | | 28.66 (17.97) | (9.05, 105.50) | 29.47 (21.6) | (7.00, 137.5) |
| | 3 | | 28.67 (19.75) | (11.29, 105.50) | 29.32 (20.65) | (8.89, 122.00) |
| | 4 | | 28.11 (17.23) | (8.00, 82.50) | 29.09 (19.68) | (9.33, 133.00) |
| **A_TH_Std_Green** | 1 | Standard deviation of the length of through green phase on "A" approach within 5-minute interval (second). | 18.26 (12.76) | (0.00, 60.25) | 18.78 (13.9) | (0.00, 99.51) |
| | 2 | | 18.21 (11.60) | (0.00, 51.04) | 18.24 (13.27) | (0.00, 89.8) |
| | 3 | | 18.12 (12.50) | (0.00, 60.09) | 18.19 (12.3) | (0.00, 68.14) |
| | 4 | | 20.04 (14.57) | (0.00, 64.55) | 18.81 (15.05) | (0.00, 164.05) |
| **A_TH_Avg_Queue** | 1 | Average through queue length at the beginning of through green phase on "A" approach (vehicle). | 12.55 (8.97) | (2.08, 40.00) | 12.54 (8.84) | (0.8, 54.00) |
| | 2 | | 11.94 (8.59) | (2.00, 40.00) | 12.77 (9.14) | (0.00, 57.00) |
| | 3 | | 12.40 (8.84) | (1.50, 40.00) | 12.87 (9.13) | (0.33, 57.00) |
| | 4 | | 12.09 (8.71) | (2.00, 40.00) | 12.82 (9.32) | (1.27, 62.00) |
| **A_TH_Avg_Wait** | 1 | Average through maximum waiting time at the beginning of through green phase on "A" approach (vehicle). | 36.49 (25.78) | (1.29, 135.50) | 36.10 (27.68) | (0.00, 142.00) |
| | 2 | | 35.48 (24.94) | (1.67, 135.50) | 36.21 (28.37) | (0.00, 175) |
| | 3 | | 36.56 (24.12) | (0.00, 135.50) | 37.06 (28.17) | (0.00, 192.5) |
| | 4 | | 36.41 (25.49) | (4.2, 135.00) | 37.27 (29.67) | (0.00, 213.00) |
| **HourlyPrecip** | - | Hourly precipitation (1/10 inch). | 0.03 (0.10) | (0.00, 0.70) | 0.09 (0.65) | (0.00, 8.00) |
| **Visibility** | - | Visibility (mile). | 9.86 (0.68) | (5.00, 10.00) | 9.64 (1.51) | (0.00, 10.00) |
| **WeatherType** | - | Weather type: 0 for normal and 1 for adverse weather. | 0.11 (0.31) | (0.00, 1.00) | 0.09 (0.28) | (0.00, 1.00) |

*Note: due to the limitation of table content, this table only list the "A" approach data. However, the within intersection dataset including the data from four approaches.*

Table 2. Summary of Variables Descriptive Statistics for the Intersection Entrance Area (Crash and Non-crash Events)

| Variable | Time Slice | Description | Crash Events | | Non-Crash Events | |
|---|---|---|---|---|---|---|
| | | | Mean (Std) | (Min, Max) | Mean (Std) | (Min, Max) |
| **Avg_speed** | 1 | Average speed on the upstream segment of "A" approach within 5-minute interval (mph). | 25.61 (8.63) | (7.75, 42.00) | 26.77 (9.44) | (5.33, 53.50) |
| | 2 | | 27.16 (9.66) | (5, 45.17) | 26.77 (9.24) | (4.83, 56.50) |
| | 3 | | 27.24 (9.88) | (4.75, 50.33) | 27.35 (10.05) | (5.17, 55.14) |
| | 4 | | 26.94 (8.82) | (4.00, 47.67) | 27.09 (10.11) | (6.00, 57.50) |
| **Std_speed** | 1 | Speed standard deviation on the upstream segment of "A" approach within 5-minute interval (mph). | 10.49 (4.96) | (0.00, 25.36) | 11.02 (5.06) | (0.00, 28.28) |
| | 2 | | 11.28 (5.04) | (0.53, 24.02) | 11.03 (5.14) | (0.58, 31.11) |
| | 3 | | 11.06 (4.9) | (0.96, 24.75) | 10.61 (4.81) | (0.58, 25.46) |
| | 4 | | 11.72 (4.75) | (0.00, 23.83) | 10.86 (4.88) | (0.00, 29.70) |
| **A_Vol_LT** | 1 | Left turn volume of "A" approach within 5-minute interval (vehicle). | 16.46 (12.31) | (0.00, 55.67) | 18.30 (13.68) | (0.00, 101.33) |
| | 2 | | 16.24 (12.00) | (0.00, 55.67) | 18.47 (13.81) | (0.00, 101.33) |
| | 3 | | 16.14 (11.54) | (0.00, 55.67) | 18.09 (13.41) | (0.00, 92.93) |
| | 4 | | 15.74 (10.36) | (0.00, 46.00) | 17.84 (13.08) | (0.00, 80.33) |
| **A_Vol_Th** | 1 | Through volume of "A" approach within 5-minute interval (vehicle). | 108.67 (64.00) | (0.00, 343.33) | 107.18 (62.64) | (0.00, 614.33) |
| | 2 | | 108.49 (63.69) | (0.00, 343.33) | 107.28 (61.21) | (0.00, 614.33) |
| | 3 | | 108.28 (64.13) | (0.00, 309.53) | 106.2 (55.34) | (0.00, 360.00) |
| | 4 | | 108.1 (63.94) | (0.00, 328.33) | 105.98 (55.48) | (0.00, 360.00) |
| **A_OAFR** | 1 | Overall average flow ratio of "A" approach within 5-minute interval. | 1.74 (3.23) | (0.95, 21.56) | 1.61 (3.26) | (0.94, 36.04) |
| | 2 | | 1.94 (3.63) | (0.95, 21.56) | 1.64 (3.32) | (0.94, 32.95) |
| | 3 | | 1.77 (3.07) | (0.95, 21.56) | 1.65 (3.3) | (0.94, 32.95) |
| | 4 | | 1.51 (2.03) | (0.95, 16.68) | 1.65 (3.67) | (0.95, 43.45) |
| **A_LT_GreenRatio** | 1 | Ratio of left turn green time on "A" approach within 5-minute interval (%). | 12.80 (6.78) | (2.67, 33.33) | 12.67 (6.32) | (1.67, 35.67) |
| | 2 | | 12.37 (6.02) | (2.67, 30.67) | 12.24 (6.08) | (2.33, 39.00) |
| | 3 | | 13.06 (6.58) | (3.33, 36.33) | 12.97 (6.74) | (1.67, 36.67) |
| | 4 | | 11.77 (5.73) | (0.33, 26.00) | 12.69 (6.67) | (1.00, 35.33) |
| **A_LT_Avg_Green** | 1 | Average length of left turn green phase on "A" approach within 5-minute interval (second). | 18.41 (9.44) | (4.00, 44.50) | 18.27 (9.42) | (3.00, 68.00) |
| | 2 | | 15.88 (7.53) | (5.00, 39.00) | 17.46 (8.07) | (4.50, 50.00) |
| | 3 | | 17.82 (9.58) | (5.00, 60.00) | 18.02 (8.75) | (5.00, 48.50) |
| | 4 | | 16.39 (7.76) | (1.00, 39.00) | 18.21 (8.91) | (3.00, 46.00) |
| **A_LT_Std_Green** | 1 | Standard deviation of the length of left turn green phase on "A" approach within 5-minute interval (second). | 4.17 (4.86) | (0.00, 18.50) | 4.92 (4.86) | (0.00, 33.94) |
| | 2 | | 5.54 (5.02) | (0.00, 23.52) | 6.17 (6.40) | (0.00, 31.82) |
| | 3 | | 5.61 (5.41) | (0.00, 24.04) | 5.47 (5.11) | (0.00, 24.75) |
| | 4 | | 4.92 (4.90) | (0.00, 23.33) | 5.09 (5.10) | (0.00, 22.63) |
| **A_LT_Avg_Queue** | 1 | Average left turn queue length at the beginning of left turn green phase on "A" approach (vehicle). | 8.53 (6.03) | (1.00, 31.00) | 8.69 (6.51) | (0.00, 37.00) |
| | 2 | | 7.92 (5.74) | (0.33, 31.00) | 8.73 (6.71) | (0.00, 37.00) |
| | 3 | | 8.00 (6.30) | (0.75, 35.00) | 8.18 (6.23) | (0.00, 37.50) |
| | 4 | | 7.81 (5.64) | (1.00, 34.00) | 8.09 (6.28) | (0.00, 38.00) |
| **A_LT_Avg_Wait** | 1 | Average left turn maximum waiting time at the beginning of left turn green phase on "A" approach (vehicle). | 102.36 (45.84) | (10.00, 178.00) | 101.86 (46.77) | (0.00, 291.00) |
| | 2 | | 95.15 (46.45) | (4.67, 169.00) | 103.11 (45.32) | (0.00, 185.5) |
| | 3 | | 96.13 (45.44) | (4.25, 188.00) | 101.72 (42.78) | (0.00, 202.00) |
| | 4 | | 97.65 (43.31) | (5.50, 170.50) | 100.31 (43.37) | (0.00, 182.00) |
| **A_TH_GreenRatio** | 1 | Ratio of through green time on "A" approach within 5-minute interval (%). | 43.92 (17.14) | (8.00, 84) | 43.64 (17.75) | (11.00, 89.33) |
| | 2 | | 42.78 (17.93) | (12.00, 80.33) | 43.88 (19.99) | (8.00, 100.00) |
| | 3 | | 44.41 (18.61) | (15.33, 83.00) | 44.31 (18.92) | (9.33, 93.00) |
| | 4 | | 45.15 (18.3) | (8.33, 87.33) | 44.51 (18.45) | (11.33, 85.67) |
| **A_TH_Avg_Green** | 1 | Average length of through green phase on "A" approach within 5-minute interval (second). | 31.24 (22.11) | (10.13, 126.00) | 29.57 (19.91) | (9.13, 134.00) |
| | 2 | | 30.71 (19.41) | (12.00, 105.50) | 31.07 (24.07) | (7.00, 137.5) |
| | 3 | | 29.49 (21.13) | (11.3, 123.00) | 28.93 (17.72) | (8.75, 132.00) |
| | 4 | | 29.32 (20.97) | (9.29, 131.00) | 30.28 (21.35) | (10.00, 126.00) |
| **A_TH_Std_Green** | 1 | Standard deviation of the length of through green phase on "A" approach within 5-minute interval (second). | 21.76 (21.10) | (0.00, 164.05) | 21.04 (15.51) | (0.00, 148.49) |
| | 2 | | 21.74 (14.01) | (0.00, 63.52) | 22.77 (21.03) | (0.00, 183.14) |
| | 3 | | 18.84 (14.12) | (0.00, 63.02) | 20.30 (15.06) | (0.00, 74.08) |
| | 4 | | 18.57 (13.67) | (0.71, 58.29) | 20.65 (15.73) | (0.00, 128.69) |
| **A_TH_Avg_Queue** | 1 | Average through queue length at the beginning of through green phase on "A" approach (vehicle). | 15.41 (10.82) | (2.00, 72.00) | 12.78 (9.77) | (1.33, 99.00) |
| | 2 | | 14.84 (10.49) | (0.00, 72.00) | 12.83 (9.91) | (1.33, 99.00) |
| | 3 | | 13.60 (10.38) | (2.50, 74.50) | 12.56 (9.87) | (1.31, 99.00) |

| | 4 | | 13.08 (10.43) | (1.45, 77.00) | 12.61 (9.95) | (1.43, 99.00) |
|---|---|---|---|---|---|---|
| **A_TH_Avg_Wait** | 1 | Average through maximum waiting time at the beginning of through green phase on "A" approach (vehicle). | 41.74 (31.75) | (1.60, 155.00) | 38.92 (29.42) | (0.00, 140.00) |
| | 2 | | 44.37 (33.46) | (0.00, 144.00) | 39.10 (31.11) | (0.00, 171.00) |
| | 3 | | 39 (32.30) | (2.25, 143.00) | 38.53 (30.57) | (0.50, 156.00) |
| | 4 | | 37.38 (32.27) | (0.33, 148.50) | 38.03 (30.30) | (0.00, 156.00) |
| **HourlyPrecip** | - | Hourly precipitation (1/10 inch). | 0.06 (0.41) | (0.00, 3.70) | 0.11 (0.71) | (0.00, 6.90) |
| **Visibility** | - | Visibility (mile). | 9.76 (0.92) | (5.00, 10.00) | 9.62 (1.56) | (0.00, 10.00) |
| **WeatherType** | - | Weather type: 0 for normal and 1 for adverse weather. | 0.09 (0.29) | (0.00, 1.00) | 0.10 (0.30) | (0.00, 1.00) |

In order to achieve a preliminary understanding about the difference between crash and non-crash events, the variable of average speed was selected as an example and the probability density distributions were presented in Figure 7 (within intersection) and Figure 8 (intersection entrance). Both Figure 7 and Figure 8 indicate that the distribution of average speed before crash events are more likely to be wide-spread than non-crash events, especially during the 5-10 minute interval. This means that the traffic condition before crash event tends to be more diverse than non-crash events, which is consistent with Theofilatos *et al.* (2018).

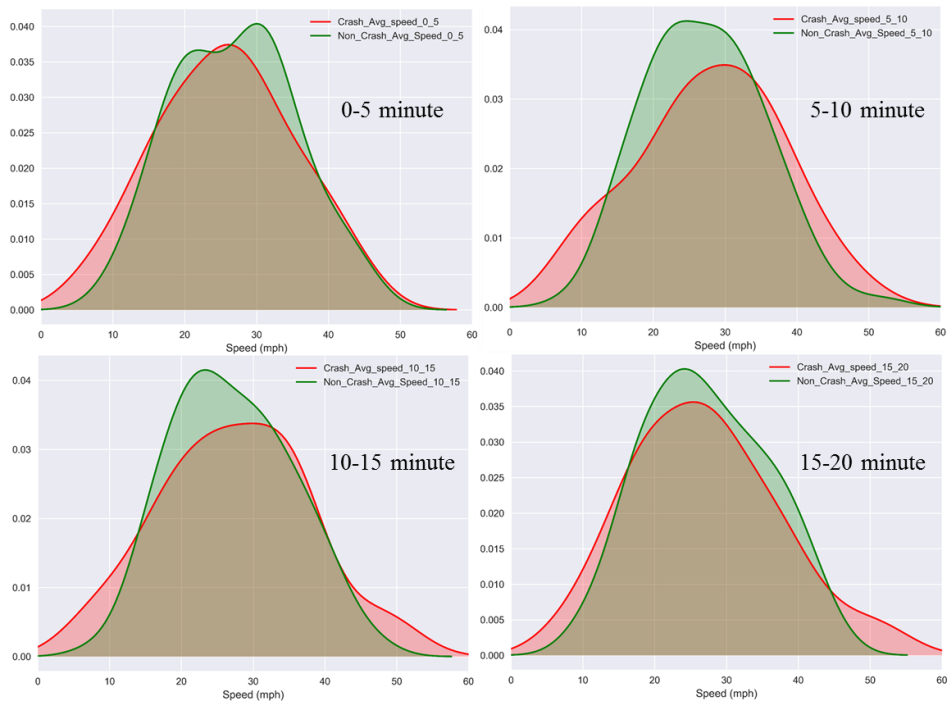

Figure 7. Distribution of the Average Speed between Crash and Non-Crash Events among Four Time Slices (Within Intersection).

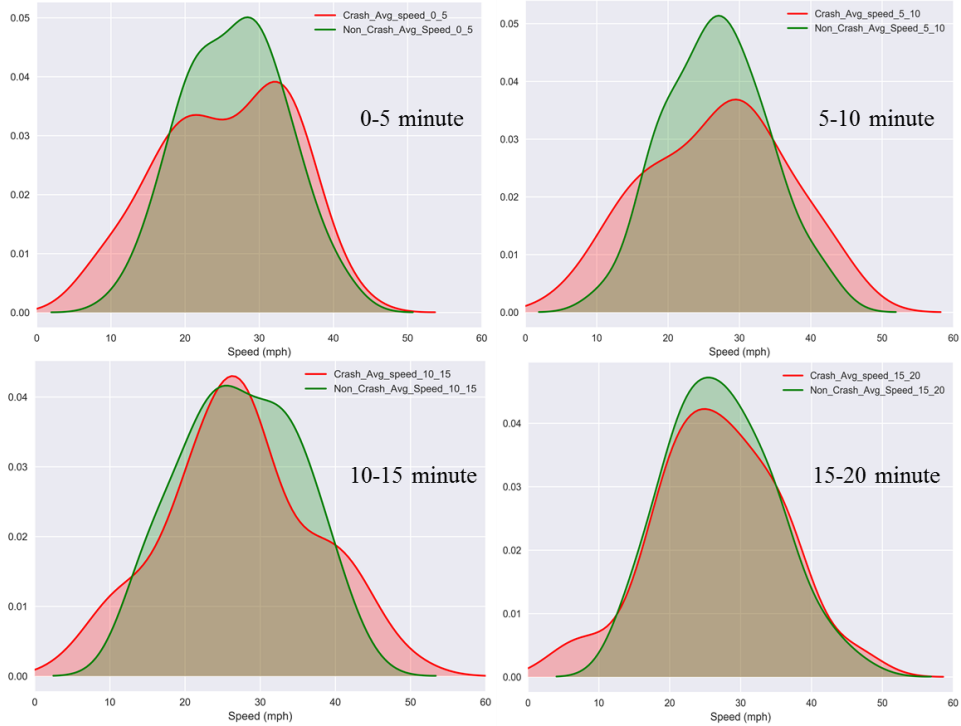

Figure 8. Distribution of the Average Speed between Crash and Non-Crash Events among Four Time Slices (Intersection Entrance).

Since the independent variables collected from different approaches are highly interactive, it is very likely that some of the independent variables are highly correlated. The threshold of 0.6 was utilized for the linear Pearson correlation analysis to identify the highly-correlated variables, which is in line with previous research (Kobelo *et al.* 2008). Moreover, with respect to the nonlinear correlation, one of the mutual information based measures, maximal information coefficient (MIC) was also employed to identify the nonlinear association between two variables (Albanese *et al.* 2018). As suggested by Albanese *et al.* (2018), the threshold of MIC was chosen to be 0.7. Above all, the highly correlated pairs of variables were selected based on two criteria: the Pearson correlation coefficient is greater than 0.6 or the MIC is greater than 0.7.

Take the time slice 1 dataset for the within intersection crashes as an example, there are 57 independent variables, which could result in 1596 ($\frac{57!}{2!(57-2)!}$) pairs of variables. The results of correlation analysis indicate that 45 pairs of highly-correlated variables were identified and presented in Table 3.

Table 3. The Highly-Correlated Variables for the Within Intersection Dataset (Time Slice 1)

| Variables | | Pearson Correlation Coefficient | Maximal Information Coefficient (MIC) |
|---|---|---|---|
| A_LT_GreenRatio_0_5 | A_LT_Avg_Green_0_5 | 0.736663 | 0.480014 |
| A_TH_GreenRatio_0_5 | A_TH_Std_Green_0_5 | 0.610484 | 0.37036 |
| A_TH_GreenRatio_0_5 | B_LT_Avg_Wait_0_5 | -0.618134 | 0.433272 |
| A_TH_GreenRatio_0_5 | C_TH_GreenRatio_0_5 | 0.718776 | 0.384575 |
| A_TH_GreenRatio_0_5 | C_TH_Avg_Green_0_5 | 0.622481 | 0.312943 |
| A_TH_GreenRatio_0_5 | D_LT_Avg_Wait_0_5 | -0.60694 | 0.399255 |
| A_TH_GreenRatio_0_5 | D_TH_GreenRatio_0_5 | -0.607984 | 0.333878 |
| A_TH_Avg_Green_0_5 | C_TH_GreenRatio_0_5 | 0.676753 | 0.398096 |
| A_TH_Avg_Green_0_5 | C_TH_Avg_Green_0_5 | 0.705255 | 0.534972 |
| A_TH_Avg_Green_0_5 | C_TH_Std_Green_0_5 | 0.673996 | 0.533665 |
| A_TH_Std_Green_0_5 | C_TH_GreenRatio_0_5 | 0.65504 | 0.494755 |
| A_TH_Std_Green_0_5 | C_TH_Avg_Green_0_5 | 0.608811 | 0.479297 |
| A_TH_Std_Green_0_5 | C_TH_Std_Green_0_5 | 0.7133 | 0.538673 |
| A_TH_Avg_Queue_0_5 | A_TH_Avg_Wait_0_5 | 0.6426 | 0.364688 |
| B_Vol_LT_0_5 | B_LT_GreenRatio_0_5 | 0.581439 | 0.709964 |
| B_Vol_LT_0_5 | B_LT_Avg_Green_0_5 | 0.617656 | 0.709964 |
| B_Vol_LT_0_5 | B_LT_Avg_Queue_0_5 | 0.537864 | 0.743575 |
| B_Vol_LT_0_5 | B_LT_Avg_Wait_0_5 | 0.443236 | 0.727099 |
| B_Vol_Th_0_5 | D_LT_Avg_Wait_0_5 | 0.612108 | 0.421989 |
| B_LT_GreenRatio_0_5 | B_LT_Avg_Green_0_5 | 0.887087 | 0.974562 |
| B_LT_GreenRatio_0_5 | B_LT_Std_Green_0_5 | 0.692994 | 0.746113 |
| B_LT_GreenRatio_0_5 | B_LT_Avg_Queue_0_5 | 0.629947 | 0.957971 |
| B_LT_GreenRatio_0_5 | B_LT_Avg_Wait_0_5 | 0.596116 | 0.957971 |
| B_LT_Avg_Green_0_5 | B_LT_Std_Green_0_5 | 0.553333 | 0.716639 |
| B_LT_Avg_Green_0_5 | B_LT_Avg_Queue_0_5 | 0.632579 | 0.960151 |
| B_LT_Avg_Green_0_5 | B_LT_Avg_Wait_0_5 | 0.752837 | 0.960151 |
| B_LT_Avg_Queue_0_5 | B_LT_Avg_Wait_0_5 | 0.574885 | 0.976834 |
| B_LT_Avg_Queue_0_5 | D_LT_Avg_Queue_0_5 | 0.325929 | 0.71544 |
| B_LT_Avg_Queue_0_5 | D_LT_Avg_Wait_0_5 | 0.406182 | 0.71544 |
| B_LT_Avg_Wait_0_5 | D_LT_Avg_Queue_0_5 | 0.400766 | 0.708458 |
| B_LT_Avg_Wait_0_5 | D_LT_Avg_Wait_0_5 | 0.649914 | 0.669127 |
| B_TH_GreenRatio_0_5 | D_TH_GreenRatio_0_5 | 0.831784 | 0.625065 |
| C_LT_GreenRatio_0_5 | C_LT_Avg_Green_0_5 | 0.721476 | 0.557538 |
| C_TH_GreenRatio_0_5 | C_TH_Std_Green_0_5 | 0.695453 | 0.432222 |
| D_Vol_LT_0_5 | D_LT_GreenRatio_0_5 | 0.618642 | 0.837144 |
| D_Vol_LT_0_5 | D_LT_Avg_Green_0_5 | 0.634303 | 0.838908 |
| D_Vol_LT_0_5 | D_LT_Avg_Queue_0_5 | 0.68267 | 0.890338 |
| D_Vol_LT_0_5 | D_LT_Avg_Wait_0_5 | 0.521591 | 0.887042 |
| D_LT_GreenRatio_0_5 | D_LT_Avg_Green_0_5 | 0.821751 | 0.965754 |
| D_LT_GreenRatio_0_5 | D_LT_Std_Green_0_5 | 0.69115 | 0.728799 |
| D_LT_GreenRatio_0_5 | D_LT_Avg_Queue_0_5 | 0.657098 | 0.922204 |
| D_LT_GreenRatio_0_5 | D_LT_Avg_Wait_0_5 | 0.601375 | 0.922204 |
| D_LT_Avg_Green_0_5 | D_LT_Avg_Queue_0_5 | 0.567144 | 0.913977 |
| D_LT_Avg_Green_0_5 | D_LT_Avg_Wait_0_5 | 0.788658 | 0.913977 |
| D_LT_Avg_Queue_0_5 | D_LT_Avg_Wait_0_5 | 0.57161 | 0.975712 |

With respect to those pairs of variables which have higher nonlinear correlation coefficients (MIC) but lower linear Pearson correlation coefficients (rows marked in grey in Table 3), a scatterplot matrix was generated to further illustrate the nonlinear association between those pairs of variables (Figure 9).

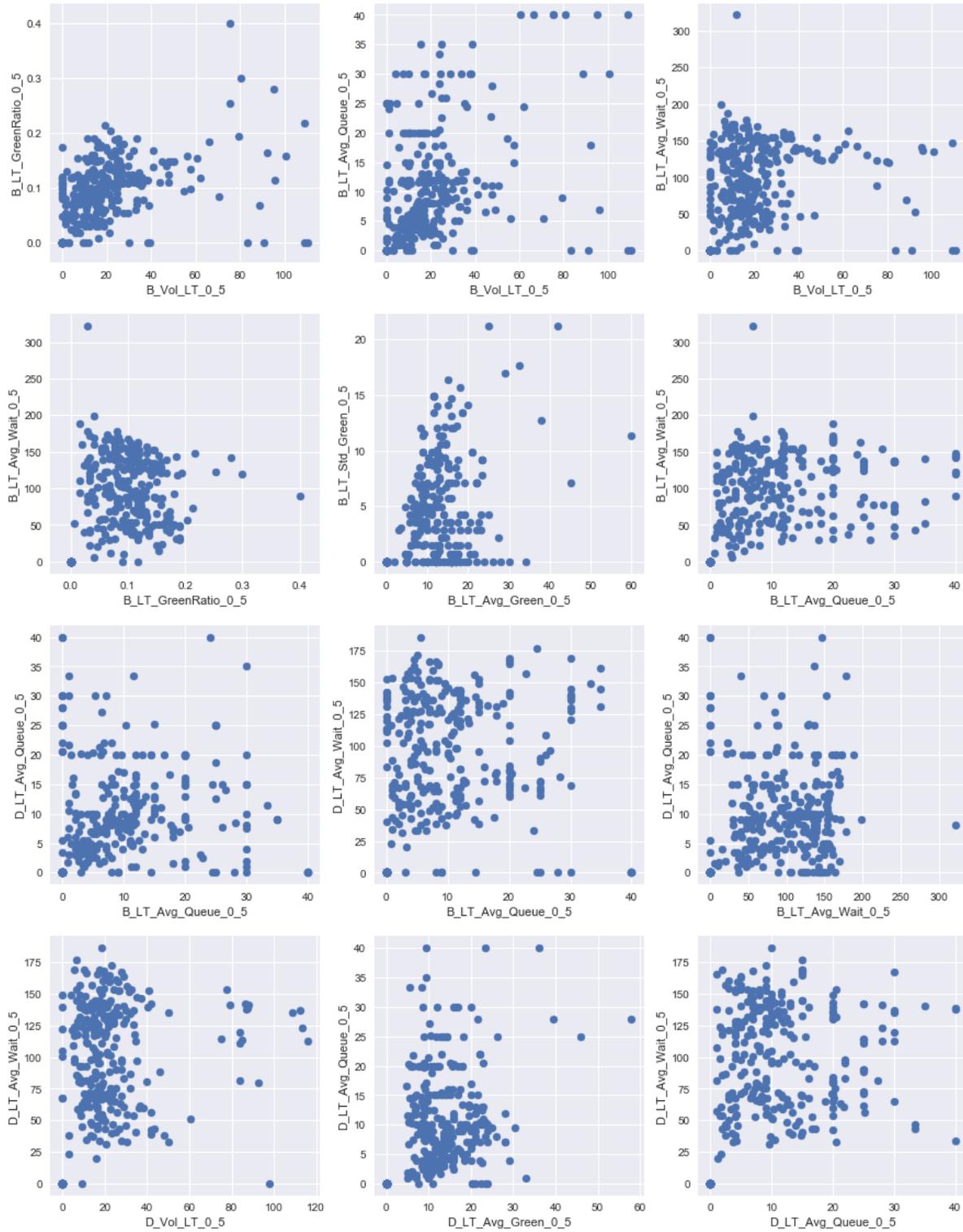

Figure 9. Scatterplot Matrix for those Variables which are Nonlinear Associated

## 3. Methodology

Suppose that there are N strata with 1 crash ($y_{ij}=1$) and m non-crash cases ($y_{ij}=0$) in stratum i, i=1, 2, …, N. Let $p_{ij}$ be the probability that the *j*th observation in the *i*th stratum is a crash; j=0, 1, 2, …, m. This crash probability could be expressed as:

$$y_{ij} \sim Bernoulli(p_{ij}) \qquad (2)$$

$$logit(p_{ij}) = \alpha_i + \beta_1 X_{1ij} + \beta_2 X_{2ij} + \cdots + \beta_k X_{kij} \qquad (3)$$

Where $\alpha_i$ is the intercept term for the *i*th stratum; $\boldsymbol{\beta} = (\beta_1, \beta_2, \ldots, \beta_k)$ is the vector of regression coefficients for *k* independent variables.

In order to take the stratification in the analysis of the observed data, the stratum-specific intercept $\alpha_i$ is considered to be nuisance parameters, and the conditional likelihood for the *i*th stratum would be expressed as (Hosmer Jr *et al.* 2013):

$$l_i(\boldsymbol{\beta}) = \frac{\exp(\sum_{u=1}^{k} \beta_u X_{ui0})}{\sum_{j=0}^{m} \exp(\sum_{u=1}^{k} \beta_u X_{uij})} \qquad (4)$$

And the full conditional likelihood is the product of the $l_i(\beta)$ over *N* strata,

$$L(\boldsymbol{\beta}) = \prod_{i=1}^{N} l_i(\boldsymbol{\beta}) \qquad (5)$$

Since the full conditional likelihood is independent of stratum-specific intercept $\alpha_i$, thus Eq. (3) cannot be used to estimate the crash probabilities. However, the estimated $\boldsymbol{\beta}$ coefficients are the log-odd ratios of corresponding variables and can be used to approximate the relative risk of an event. Furthermore, the log-odds ratios can also be used to develop a prediction model under this matched case-control analysis. Suppose two observation vectors $\boldsymbol{X_{i1}} = (X_{1i1}, X_{2i1}, \ldots, X_{Ki1})$ and $\boldsymbol{X_{i2}} = (X_{1i2}, X_{2i2}, \ldots, X_{Ki2})$ from the *i*th strata, the odds ratio of crash occurrence caused by observation vector $\boldsymbol{X_{i1}}$ relative to observation vector $\boldsymbol{X_{i2}}$ could be calculated as:

$$\frac{p_{i1}/(1-p_{i1})}{p_{i2}/(1-p_{i2})} = \exp[\sum_{k=1}^{K} \beta_k (X_{ki1} - X_{ki2})] \qquad (6)$$

The right hand side of Eq. (6) is independent of $\alpha_i$ and can be calculated using the estimated $\boldsymbol{\beta}$ coefficients. Thus, the above relative odds ratio could be utilized for predicting crash occurrences by replacing $\boldsymbol{X_{i2}}$ with the vector of the independent variables in the *i*th stratum of non-crash events. One may use simple average of each variable for all non-crash observations within the stratum. Let $\boldsymbol{\bar{X}_i} = (\bar{X}_{1i}, \bar{X}_{2i}, \ldots, \bar{X}_{Ki})$ denote the vector of mean values of non-crash events of the *k* variables within the *i*th stratum. Then the odds ratio of a crash relative to the non-crash events in the *i*th stratum could be approximated by:

$$\frac{p_{i1}/(1-p_{i1})}{p_{\bar{i}}/(1-p_{\bar{i}})} = \exp[\sum_{k=1}^{K} \beta_k (X_{ki1} - \bar{X}_{ki})] \qquad (7)$$

Full Bayesian inference was employed in this study. For each model, three chains of 20,000 iterations were set up in WinBUGS (Lunn *et al.* 2000), the first 5,000 iterations were excluded as burn-in, the latter 15,000 stored iterations were set to estimate the posterior distribution. Convergence was evaluated using the built-in Brooks-Gelman-Rubin (BGR) diagnostic statistic (Brooks and Gelman 1998).

In terms of model goodness-of-fit, the AUC value which is the area under Receiver Operating Characteristic (ROC) curve was also adopted. The ROC curve illustrates the relationship between the true positive rate (sensitivity) and the false alarm rate (1–specificity) of model classification results based on a given threshold from 0 to 1. It is worth noting that the classification results of Bayesian random parameters logistic model is based on the predicted crash probabilities, which lie in the range of 0 to 1, while the classification result of Bayesian conditional logistic model and Bayesian random parameters conditional logistic model are based on the predicted odds ratio, which may be larger than 1. In order to be consistent with the other two models, all the odds ratios predicted by Bayesian conditional logistic model were divided by the maximum odds ratio to create adjusted odds ratios. Later on, the adjusted odds ratios were used to create the classification result based on different threshold from 0 to 1. In this study, AUC values were calculated using R package pROC (Robin *et al.* 2011).

## 4. Model results

### 4.1. Within intersection crashes

This section discusses the modeling results of the Bayesian conditional logistic models for the within intersection crashes based on the full dataset (four time slices) and different time slices datasets, respectively. Table 4 shows the results of within intersection model based on full dataset. In total, 14 variables were identified to be significant variables, including speed characteristics, signal timing, queue length, and waiting time related factors collected from different approaches and time slices.

Table 4. Results of the Bayesian Conditional Logistic Model based on Full Dataset (Within Intersection).

| Variables | Coefficient Estimation | | Odds Ratio | |
| --- | --- | --- | --- | --- |
| | Mean | 95% BCI | Mean | 95% BCI |
| Avg_speed_0_5 | **-0.038** | **(-0.07, -0.005)*** | **0.963** | **(0.932, 0.995)*** |
| Std_speed_0_5 | **0.066** | **(0.001, 0.131)** | **1.068** | **(1.001, 1.14)** |
| B_TH_Avg_Wait_0_5 | **0.013** | **(0.002, 0.024)** | **1.013** | **(1.002, 1.024)** |
| D_TH_Avg_Wait_0_5 | **0.016** | **(0.006, 0.026)** | **1.016** | **(1.006, 1.026)** |
| B_LT_Std_Green_5_10 | **-0.138** | **(-0.248, -0.04)** | **0.871** | **(0.78, 0.961)** |
| C_TH_Avg_Wait_5_10 | **0.017** | **(0.001, 0.032)*** | **1.017** | **(1.001, 1.033)*** |
| B_Vol_LT_10_15 | **0.029** | **(0.005, 0.054)*** | **1.029** | **(1.005, 1.055)*** |
| D_TH_Avg_Green_10_15 | **-0.059** | **(-0.103, -0.017)** | **0.943** | **(0.902, 0.983)** |
| A_LT_Avg_Green_15_20 | **-0.055** | **(-0.106, -0.006)** | **0.946** | **(0.899, 0.994)** |
| A_LT_Std_Green_15_20 | **-0.090** | **(-0.161, -0.019)** | **0.914** | **(0.851, 0.981)** |
| C_LT_Avg_Queue_15_20 | **-0.094** | **(-0.18, -0.013)** | **0.910** | **(0.835, 0.987)** |
| D_TH_GreenRatio_15_20 | **-0.088** | **(-0.175, -0.004)** | **0.916** | **(0.839, 0.996)** |
| D_TH_Std_Green_15_20 | **0.060** | **(0.004, 0.114)** | **1.062** | **(1.004, 1.121)** |
| D_TH_Avg_Queue_15_20 | **-0.067** | **(-0.13, -0.005)** | **0.935** | **(0.878, 0.995)** |
| AUC | 0.7596 | | | |

*Note: 95% BCI values marked in bold and noted by * indicate that these variables are significant at the 0.1 level, while other variables are significant at the 0.05 level.*

Considering that the traffic and signal characteristics during different time slice may have different relationship with the real-time crash risk. To investigate the differences between different time-slice datasets, four separate time-slice models were developed based on four time slices, respectively. Table 5 shows the results of 4 time-slice models for the within intersection dataset. The model comparison results based on AUC values indicate that the slice 2 model performs the best, followed by the slice 4 and slice 1 models. However, based on slice 1 model, there would be no spare time to implement any proactive traffic management strategy to prevent the possibility of crash occurrence. Moreover, as stated by Golob et al. (2004), there may exist 2.5 min difference between the exact crash time and reported crash time, thus the slice 1 model was treated as a reference. Finally, the slice 2 model was selected to conduct further interpretation.

Table 5. Results of the Bayesian Conditional Logistic Models based on Different Time Slices (Within Intersection).

| Variables | Slice 1 | | Slice 2 | | Slice 3 | | Slice 4 | |
|---|---|---|---|---|---|---|---|---|
| | Mean (95% BCI) | Odds Ratio (95% BCI) | Mean (95% BCI) | Odds Ratio (95% BCI) | Mean (95% BCI) | Odds Ratio (95% BCI) | Mean (95% BCI) | Odds Ratio (95% BCI) |
| Avg_speed | **-0.033 (-0.063, -0.004)*** | **0.968 (0.939, 0.996)*** | - | - | - | - | - | - |
| Std_speed | **0.056 (0.008, 0.101)*** | **1.058 (1.008, 1.106)*** | - | - | - | - | - | - |
| A_Vol_Th | - | - | **0.005 (0.001, 0.011)*** | **1.005 (1.001, 1.011)*** | - | - | - | - |
| A_LT_Avg_Green | - | - | - | - | - | - | **-0.041 (-0.08, -0.003)*** | **0.96 (0.923, 0.997)*** |
| A_LT_Std_Green | - | - | - | - | - | - | **-0.064 (-0.131, -0.004)** | **0.938 (0.877, 0.996)** |
| B_Vol_LT | **0.034 (0.009, 0.063)** | **1.035 (1.009, 1.065)** | **0.039 (0.011, 0.07)** | **1.040 (1.011, 1.073)** | **0.031 (0.005, 0.058)** | **1.031 (1.005, 1.06)** | **0.036 (0.006, 0.066)** | **1.037 (1.006, 1.068)** |
| B_LT_Std_Green | - | - | **-0.106 (-0.206, -0.017)** | **0.899 (0.814, 0.983)** | - | - | - | - |
| B_TH_Avg_Queue | - | - | **-0.046 (-0.09, -0.005)*** | **0.955 (0.914, 0.995)*** | - | - | **-0.052 (-0.103, -0.008)** | **0.949 (0.902, 0.992)** |
| B_TH_Avg_Wait | **0.013 (0.003, 0.022)** | **1.013 (1.003, 1.022)** | - | - | - | - | - | - |
| C_Vol_Th | **-0.006 (-0.012, 0.000)*** | **0.994 (0.988, 1.000)*** | - | - | - | - | - | - |
| C_LT_Avg_Queue | - | - | - | - | - | - | **-0.076 (-0.159, -0.003)** | **0.927 (0.853, 0.997)** |
| D_Vol_LT | - | - | **-0.036 (-0.067, -0.004)*** | **0.965 (0.935, 0.996)*** | - | - | **-0.039 (-0.078, -0.004)** | **0.962 (0.925, 0.996)** |
| D_OAFR | - | - | **0.518 (0.077, 0.978)** | **1.679 (1.08, 2.659)** | - | - | - | - |
| D_TH_GreenRatio | - | - | - | - | - | - | **-0.074 (-0.145, -0.004)** | **0.929 (0.865, 0.996)** |
| D_TH_Avg_Green | - | - | - | - | **-0.057 (-0.099, -0.019)** | **0.945 (0.906, 0.981)** | - | - |
| D_TH_Std_Green | - | - | - | - | - | - | **0.054 (0.006, 0.103)** | **1.055 (1.006, 1.108)** |
| D_TH_Avg_Wait | **0.009 (0.000, 0.017)** | **1.009 (1.000, 1.017)** | **-0.011 (-0.02, -0.002)** | **0.989 (0.98, 0.998)** | **0.011 (0.001, 0.021)** | **1.011 (1.001, 1.021)** | - | - |
| AUC | 0.6759 | | 0.6927 | | 0.6337 | | 0.6858 | |

*Note: Mean (95% BCI) values marked in bold are significant at the 0.05 level; Mean (95% BCI) values marked in bold and noted by * are significant at the 0.1 level.*

It is worth noting that the speed related variables were only found to be significant in slice 1 model, which might be explained as that the speed characteristics on the upstream segment only have short-term impacts on the within intersection crash occurrence, and relatively, these within intersection crashes are more likely to be influenced by the signal timing and traffic volume related variables. Based on the estimation results of slice 2 model, seven variables were found to be significantly associated with the crash risk within intersection area: (1) The positive coefficient (0.005) of "A_Vol_Th" indicates that higher through volume from "A" approach tends to increase the crash risk, which is consistent with previous aggregated intersection studies (Poch and Mannering 1996, Chin and Quddus 2003, Abdel-Aty and Wang 2006, Guo *et al.* 2010) that higher exposure may results in more crashes. The odds ratio of 1.005 means that when other variables held constant, one-unit increase in the through volume from "A" approach would increase the odds of crash occurrence by 0.5%; (2) Similarly, the left turn volume from "B" approach was also found to be positively correlated with the odds of crash occurrence. This could be explained in that higher left turn volume from "B" approach may results in more conflicts between the through vehicles from "A" approach and the left turn vehicle from "B" approach. The odds ratio of 1.04 means that when other variables held constant, one-unit increase in the left turn volume from "B" approach would increase the odds of crash occurrence by 4%; (3) "B_LT_Std_Green" was found to be negatively associated with the odds of crash occurrence within intersection, which means that higher standard deviation of the length of left turn phase on "B" approach could improve the safety performance of intersection. The possible reason is that when the left turn volume from "B" approach, as well as other variables held constant, the higher variation in the length of left turn phase on "B" approach indicates higher adaptability of the left turn phase, which indeed increase the safety performance of intersection; (4) "B_TH_Avg_Queue" was found to have negative effect on the crash risk within intersection, which could be explained as that higher queue length on the through lanes of "B" approach may represent that more signal priority has been given to the "A" approach, which may reduce the exposed conflicting traffic flows between through vehicles from "A" and "B" approaches; (5) The negative coefficient (-0.036) of "D_Vol_LT" indicates that higher left turn volume from "D" approach tends to reduce the crash risk within intersection. The possible reason might be that more left turn vehicle from "D" approach may raise the awareness of those drivers from the "A" approach, which will therefore reduce the odds of crash occurrence. This is similar to the previous findings by Guo *et al.* (2010), which indicated that the left-turn ADT on minor road is negatively associated with the crash frequency at signalized intersections; (6) Higher "D_OAFR" tends to increase the odds of crash occurrence, which demonstrates that higher variation in traffic flow across through lanes on "D" approach tends to increase the crash risk within intersection. This could be potentially explained by that higher variation in traffic flow across through lanes on "D" approach may results in many lane change behavior occurring within the intersection, which will increase the complexity of traffic flow within intersection, as well as the odds of crash occurrence within intersection; (7) "D_TH_Avg_Wait" was found to be negatively correlated with the odds of crash occurrence within the intersection. This might be explained by that a longer waiting time on "D" approach indicates higher signal priority was given to the "A" approach, which will indeed reduce the exposed conflicting traffic flows between the through vehicles from "A" and "D" approaches.

*4.2. Intersection entrance crashes*

Similar to the within intersection crashes, a full model was first developed for the intersection entrance crashes based on four time slices. Table 6 shows the results of intersection entrance model based on full dataset. In total, 7 variables were identified to be significant variables, including speed characteristics, signal timing, queue length, and waiting time related factors collected from different time slices.

Table 6. Results of the Bayesian Conditional Logistic Model based on Full Dataset (Intersection Entrance).

| Variables | Coefficient Estimation | | Odds Ratio | |
|---|---|---|---|---|
| | Mean | 95% BCI | Mean | 95% BCI |
| A_TH_Avg_Queue_0_5 | **0.054** | **(0.018, 0.094)** | **1.055** | **(1.018, 1.099)** |
| A_LT_Avg_Green_5_10 | **-0.056** | **(-0.107, -0.006)** | **0.946** | **(0.899, 0.994)** |
| A_LT_Avg_Queue_5_10 | **-0.065** | **(-0.128, -0.007)*** | **0.937** | **(0.88, 0.993)*** |
| A_TH_Avg_Wait_5_10 | **0.014** | **(0.000, 0.028)** | **1.014** | **(1.000, 1.028)** |
| Avg_speed_10_15 | **-0.046** | **(-0.078, -0.017)** | **0.955** | **(0.925, 0.983)** |
| A_TH_Avg_Green_15_20 | **-0.037** | **(-0.069, -0.009)** | **0.964** | **(0.933, 0.991)** |
| A_LT_GreenRatio_15_20 | **-0.084** | **(-0.167, -0.003)** | **0.919** | **(0.846, 0.997)** |
| AUC | 0.728 | | | |

*Note: 95% BCI values marked in bold and noted by * indicate that these variables are significant at the 0.1 level, while other variables are significant at the 0.05 level.*

In addition to the full model, four separate time-slice models were developed for the intersection entrance crashes based on four time slices, respectively. Table 7 shows the results of 4 time-slice models for the intersection entrance dataset. The model comparison results based on AUC values indicate that the slice 2 model performs the best, followed by the slice 4 and slice 1 models, which is in line with the within intersection models. The possible reason why the slice 4 model also performs very well might be that the traffic environment in the intersection entrance area is more simple than the within intersection area, therefore, the crash risk in the intersection entrance area tends to be more stable over time than the within intersection area. However, there may exist some uncertainty because of the insufficient sample size, which will afterwards influence the performance of different time-slice model. It is worth noting that the sign of the significant variables are consistent in all slices. Therefore, all the 7 significant variables among four time-slice models will be investigated for the intersection entrance dataset.

Table 7. Results of the Bayesian Conditional Logistic Models based on Different Time Slices (Intersection Entrance).

| Variables | Slice 1 | | Slice 2 | | Slice 3 | | Slice 4 | |
|---|---|---|---|---|---|---|---|---|
| | Mean (95% BCI) | Odds Ratio (95% BCI) | Mean (95% BCI) | Odds Ratio (95% BCI) | Mean (95% BCI) | Odds Ratio (95% BCI) | Mean (95% BCI) | Odds Ratio (95% BCI) |
| Avg_speed | **-0.050 (-0.077, -0.024)** | **0.951 (0.926, 0.976)** | **-0.041 (-0.072, -0.012)** | **0.96 (0.931, 0.988)** | **-0.038 (-0.066, -0.01)** | **0.963 (0.936, 0.99)** | **-0.037 (-0.068, -0.006)** | **0.964 (0.934, 0.994)** |
| A_Vol_LT | **-0.048 (-0.086, -0.013)** | **0.953 (0.918, 0.987)** | **-0.037 (-0.07, -0.005)*** | **0.964 (0.932, 0.995)*** | **-0.046 (-0.086, -0.01)** | **0.955 (0.918, 0.99)** | **-0.047 (-0.091, -0.009)** | **0.954 (0.913, 0.991)** |
| A_LT_Avg_Green | - | - | - | - | - | - | **-0.050 (-0.096, -0.003)** | **0.951 (0.908, 0.997)** |
| A_LT_Avg_Wait | - | - | **-0.013 (-0.022, -0.003)** | **0.987 (0.978, 0.997)** | - | - | - | - |
| A_TH_GreenRatio | - | - | **-0.040 (-0.081, -0.002)** | **0.961 (0.922, 0.998)** | - | - | - | - |
| A_TH_Std_Green | - | - | - | - | **-0.035 (-0.075, 0)** | **0.966 (0.928, 1)** | **-0.041 (-0.077, -0.007)** | **0.960 (0.926, 0.993)** |
| A_TH_Avg_Queue | **0.030 (0.001, 0.061)*** | **1.030 (1.001, 1.063)*** | - | - | - | - | - | - |
| AUC | 0.6679 | | 0.6770 | | 0.6466 | | 0.6767 | |

*Note: Mean (95% BCI) values marked in bold are significant at the 0.05 level; Mean (95% BCI) values marked in bold and noted by * are significant at the 0.1 level.*

In total, seven variables from the "A" approach were found to be significantly correlated with the crash occurrence in the intersection entrance area: (1) the coefficients of average speed are consistent to be negative among four time-slice models, which means that lower average speed tends to increase the odds of crash occurrence in the intersection entrance area, which is consistent with previous studies (Abdel-Aty *et al.* 2012, Ahmed *et al.* 2012a, b, Ahmed and Abdel-Aty 2012, Xu *et al.* 2012, Shi and Abdel-Aty 2015, Yu *et al.* 2016, Yuan *et al.* 2018). This could be explained by that the lower average speed, i.e., congested condition, are more likely to have higher crash risk than uncongested condition; (2) the left turn volume was found to have significant negative effect on the odds of crash occurrence, which means that higher left turn volume may results in lower crash risk. The possible reason might be that driver intending to turn left approach the intersection more carefully and with lower speeds. Thus higher left turn volume may increase the driver awareness when approaching the entering approach, which may improve the safety performance; (3) the average length of left turn green phase was found to be negatively correlated with the odds of crash occurrence, which means that when the left turn volume, as well as other variables held constant, longer left turn green time could decrease the odds of crash occurrence; (4) the negative coefficient of the left turn average waiting time demonstrates that the longer waiting time for the left turn vehicles may results in better safety performance. The possible reason might be that the longer waiting time for the left turn vehicles, the less exposure may exist between left turn and through vehicles, which may reduce the crash risk; (5) similarly, the green ratio, as well as the standard deviation of the green time of the through phase were found to have negative effect on the odds of crash occurrence, which indicate that longer and more adaptive green phase for the through vehicles could significantly improve the safety performance of the intersection entrance area. It may be reasoned that longer and more adaptive green phase for the through vehicles could significantly decrease the frequency of stop-and-go traffic, which will therefore decrease the potential conflicts. Similarly, Lee *et al.* (2013) found that the implementation of cooperative vehicle intersection control algorithm, which optimize the vehicle trajectory to reduce the stop-and-go frequency, can reduce the number of rear-end crash events by 30-87% for different volume condition; (6) the positive coefficient of average queue on the through lanes indicates that longer queue on the through lanes may increase the odds of crash occurrence.

## 5. Discussion and Conclusion

This research examined the real-time crash risk at signalized intersections based on the disaggregated data from multiple sources, including travel speed collected by Bluetooth detectors, lane-specific traffic volume and signal timing data from adaptive signal controllers, and weather data collected by airport weather station. The intersection and intersection-related crashes were collected and then divided into three types, i.e., within intersection crashes, intersection entrance crashes, and intersection exit crashes. In terms of the sample size, only the within intersection crashes and intersection entrance crashes were considered and then modeled separately. Matched case-control design with a control-to-case ratio of 4:1 was employed to select the corresponding non-crash events for each crash event, and three confounding factors, i.e., location, time of day, and day of the week, were selected as matching factors. Afterwards, all the traffic, signal timing, and weather characteristics during 20-minute window prior to the crash or non-crash events were collected and divided into four 5-minute slices, i.e., 0-5 minute, 5-10 minute, 10-15 minute, and 15-20 minute. Later on, Bayesian conditional logistic models were developed for within intersection crashes and intersection entrance crashes, respectively.

For the within intersection crashes, the results of the full model (based on four time-slice datasets) indicate that 14 variables are significantly associated with the real-time crash risk, including speed characteristics, signal timing, queue length, and waiting time related factors collected from different approaches and time slices. The AUC value of the full model is 0.7596, which is much higher than the time-slice models. This comparison result reveals that incorporating all time slices variables could significantly improve the model performance. With respect to the four time-slice models, the model results show that the slice 2 model performs much better than the other modes in terms of the AUC value, which means that the characteristics during 5-10 minutes prior to the crash event have more power in the real-time crash risk prediction than the other time intervals. Among the slice 2 model, three volume related variables, i.e., the through volume from "A" approach (at-fault vehicle traveling approach), the left turn volume from "B" approach (near-side crossing approach), and the OAFR from "D" approach (far-side crossing approach), were found to have significant positive effects on the odds of crash occurrence, which is consistent with previous aggregated studies(Chin and Quddus 2003, Abdel-Aty and Wang 2006, Guo *et al.* 2010, Xie *et al.* 2013, Wang *et al.* 2016). However, the left turn volume from "D" approach was found to have negative effect on the crash risk, this may be reasoned that more left turn vehicle from "D" approach may raise the awareness of those drivers from "A" approach, which will therefore reduce the crash risk.

Moreover, the standard deviation of the length of left turn green phase of "B" approach, the average queue length of the through vehicles on "B" approach, and the average waiting time of the through vehicles on "D" approach were found to be negatively associated with the odds of crash occurrence. These findings imply that the increased adaptability for the left turn signal timing of "B" approach (higher "B_LT_Std_Green") and increased priority for "A" approach (higher "B_TH_Avg_Queue" and "D_TH_Avg_Wait") could significantly decrease the odds of crash occurrence caused by the vehicles from "A" approach. It is worth noting that the speed-related variables were only found to be significant in the slice 1 model. This might be because the potential conflicting movements within intersection area are quite dynamic, and the speed characteristics on the upstream segment may only have short-term impacts on the within intersection crash occurrence.

With respect to the intersection entrance crashes, since all the involving vehicles in the intersection entrance crash are traveling on the same approach with the at-fault vehicle, only the characteristics of "A" approach were included in the models. The full model performs much better than the four time-slice models in terms of the AUC value, which is in line with the within intersection models. Among the four time-slice models, the slice 2 model performs the best, which is slightly better than the slice 4 and slice 1 models. The possible reason why the slice 4 model also performs very well might be that the traffic environment in the intersection entrance area is more simple than the within intersection area, therefore, the crash risk in the intersection entrance area tends to be more stable over time than the within intersection area, and the insufficient sample size may also results in some uncertainty among the four time-slice models. Therefore, the significant variables in four time-slice models were investigated. Average speed was found to have significant negative effect on the odds of crash occurrence, which is consistent with previous studies (Abdel-Aty *et al.* 2012, Ahmed *et al.* 2012a, b, Ahmed and Abdel-Aty 2012, Xu *et al.* 2012, Shi and Abdel-Aty 2015, Yu *et al.* 2016, Yuan *et al.* 2018). The left turn volume was surprisingly found to be negatively correlated with the odds of crash occurrence, which might be explained as the higher left turn volume may increase the driver awareness when approaching the entering approach, which may improve the safety performance. Moreover, three signal timing variables, i.e., A_LT_Avg_Green, A_TH_GreenRatio, and A_TH_Std_Green, were found to have significant negative effects on the odds of crash occurrence. These findings imply that longer average green time for the left turn phase, higher green ratio for the through phase, and higher adaptability for the through green phase can significantly improve the safety performance in the intersection entrance area. Besides, the average queue length on the through lanes was found to have positive effect on the odds of crash occurrence, which indicates that longer queue on the through lanes may significantly increase the crash risk.

It is worth noting that all the weather related variables are insignificant in both within intersection models and intersection entrance models. This might be explained by that the weather related variables are more likely to have effects on high-speed segment or free-flow facilities, while the signalized intersections are usually operated at low speed and they are highly interrupted by the traffic signals, therefore, the weather related variables may not have significant effects on the crash occurrence at signalized intersections. Above all, the model results provide a lot of insights on the relationship between the crash risk at signalized intersection and the real-time traffic and signal timing characteristics. For example, the results related to signal timing variables imply that higher adaptability for both left turn and through phases, longer average green time for the left turn phase, and higher green ratio for the through phase could significantly improve the safety performance of signalized intersections. These findings might be incorporated into the adaptive signal control algorithm to better accommodate the real-time safety and efficiency requirements.

Overall, this study succeed in verifying the feasibility of real-time safety analysis for signalized intersections. However, there are still some limitations for the current study. For example, only 23 signalized intersections on three corridors were considered, which may results in some bias in the data collection even though the matched case-control design was utilized. Also, different geometric characteristics may also have significant effects on real-time crash risk, which has already been demonstrated by Ahmed *et al.* (2012a). However, the geometric effects were controlled in this study by using matched case-control design. Above all, further investigation would be beneficial to improve the generalization of the model results, which may start from the following aspects: increase the sample size by collecting data from large-scale signalized intersections which may also have various geometric characteristics and try to use unbalanced dataset which is more realistic than the artificially balanced data. It is also worth noting that the vulnerable users (pedestrians, motorcyclists) related crashes were not considered in the current stage, although signalized intersections are typical dangerous hotspots for the vulnerable road users. Therefore, it would be meaningful to investigate the relationship between vulnerable-user-related-crash occurrence and real-time traffic and signal characteristics.

# References

Abdel-Aty, M., Pande, A., Das, A., Knibbe, W., 2008. Assessing safety on dutch freeways with data from infrastructure-based intelligent transportation systems. Transportation Research Record: Journal of the Transportation Research Board (2083), 153-161.

Abdel-Aty, M., Uddin, N., Pande, A., Abdalla, F., Hsia, L., 2004. Predicting freeway crashes from loop detector data by matched case-control logistic regression. Transportation Research Record: Journal of the Transportation Research Board (1897), 88-95.

Abdel-Aty, M., Wang, X., 2006. Crash estimation at signalized intersections along corridors: Analyzing spatial effect and identifying significant factors. Transportation Research Record: Journal of the Transportation Research Board (1953), 98-111.

Abdel-Aty, M.A., Hassan, H.M., Ahmed, M., Al-Ghamdi, A.S., 2012. Real-time prediction of visibility related crashes. Transportation Research Part C: Emerging Technologies 24, 288-298.

Ahmed, M., Abdel-Aty, M., 2013. A data fusion framework for real-time risk assessment on freeways. Transportation Research Part C: Emerging Technologies 26, 203-213.

Ahmed, M., Abdel-Aty, M., Yu, R., 2012a. Assessment of interaction of crash occurrence, mountainous freeway geometry, real-time weather, and traffic data. Transportation Research Record: Journal of the Transportation Research Board 2280, 51-59.

Ahmed, M., Abdel-Aty, M., Yu, R., 2012b. Bayesian updating approach for real-time safety evaluation with automatic vehicle identification data. Transportation Research Record: Journal of the Transportation Research Board 2280, 60-67.

Ahmed, M.M., Abdel-Aty, M.A., 2012. The viability of using automatic vehicle identification data for real-time crash prediction. IEEE Transactions on Intelligent Transportation Systems 13 (2), 459-468.

Albanese, D., Riccadonna, S., Donati, C., Franceschi, P., 2018. A practical tool for maximal information coefficient analysis. GigaScience 7 (4), giy032.

Basso, F., Basso, L.J., Bravo, F., Pezoa, R., 2018. Real-time crash prediction in an urban expressway using disaggregated data. Transportation Research Part C: Emerging Technologies 86, 202-219.

Brooks, S.P., Gelman, A., 1998. General methods for monitoring convergence of iterative simulations. Journal of computational and graphical statistics 7 (4), 434-455.

Cai, Q., Abdel-Aty, M., Lee, J., Huang, H., 2018a. Integrating macro-and micro-level safety analyses: A bayesian approach incorporating spatial interaction. Transportmetrica A: Transport Science, 1-22.

Cai, Q., Abdel-Aty, M., Lee, J., Wang, L., Wang, X., 2018b. Developing a grouped random parameters multivariate spatial model to explore zonal effects for segment and intersection crash modeling. Analytic Methods in Accident Research 19, 1-15.

Chin, H.C., Quddus, M.A., 2003. Applying the random effect negative binomial model to examine traffic accident occurrence at signalized intersections. Accident Analysis & Prevention 35 (2), 253-259.

Chung, W., Abdel-Aty, M., Lee, J., 2018. Spatial analysis of the effective coverage of land-based weather stations for traffic crashes. Applied Geography 90, 17-27.

Dong, C., Clarke, D.B., Yan, X., Khattak, A., Huang, B., 2014. Multivariate random-parameters zero-inflated negative binomial regression model: An application to estimate crash frequencies at intersections. Accident Analysis & Prevention 70, 320-329.

Golob, T.F., Recker, W.W., Alvarez, V.M., 2004. Freeway safety as a function of traffic flow. Accident Analysis & Prevention 36 (6), 933-946.

Guo, F., Wang, X., Abdel-Aty, M.A., 2010. Modeling signalized intersection safety with corridor-level spatial correlations. Accident Analysis & Prevention 42 (1), 84-92.

Hosmer Jr, D.W., Lemeshow, S., Sturdivant, R.X., 2013. Applied logistic regression John Wiley & Sons, Hoboken, New Jersey.

Khattak, Z.H., Fontaine, M.D., Boateng, R.A., Year. Evaluating the impact of adaptive signal control technology on driver stress and behavior. In: Proceedings of the Transportation Research Board 97th Annual MeetingTransportation Research Board, Washington D.C.

Khattak, Z.H., Magalotti, M.J., Fontaine, M.D., 2018b. Estimating safety effects of adaptive signal control technology using the empirical bayes method. Journal of Safety Research 64, 121-128.

Kobelo, D., Patrangenaru, V., Mussa, R., 2008. Safety analysis of florida urban limited access highways with special focus on the influence of truck lane restriction policy. Journal of Transportation Engineering 134 (7), 297-306.

Lee, C., Abdel-Aty, M., Hsia, L., 2006. Potential real-time indicators of sideswipe crashes on freeways. Transportation Research Record: Journal of the Transportation Research Board (1953), 41-49.

Lee, C., Hellinga, B., Saccomanno, F., 2003. Real-time crash prediction model for application to crash prevention in freeway traffic. Transportation Research Record: Journal of the Transportation Research Board (1840), 67-77.

Lee, J., Abdel-Aty, M., Cai, Q., 2017. Intersection crash prediction modeling with macro-level data from various geographic units. Accident Analysis & Prevention 102, 213-226.

Lee, J., Park, B.B., Malakorn, K., So, J.J., 2013. Sustainability assessments of cooperative vehicle intersection control at an urban corridor. Transportation Research Part C: Emerging Technologies 32, 193-206.

Lunn, D.J., Thomas, A., Best, N., Spiegelhalter, D., 2000. Winbugs-a bayesian modelling framework: Concepts, structure, and extensibility. Statistics and computing 10 (4), 325-337.

Mussone, L., Bassani, M., Masci, P., 2017. Analysis of factors affecting the severity of crashes in urban road intersections. Accid Anal Prev 103, 112-122.

Oh, C., Oh, J.-S., Ritchie, S., Chang, M., 2001. Real-time estimation of freeway accident likelihood. 80th Annual Meeting of the Transportation Research Board, Washington, DC. Washington, D.C.

Poch, M., Mannering, F., 1996. Negative binomial analysis of intersection-accident frequencies. Journal of transportation engineering 122 (2), 105-113.

Robin, X., Turck, N., Hainard, A., Tiberti, N., Lisacek, F., Sanchez, J.-C., Müller, M., 2011. Proc: An open-source package for r and s+ to analyze and compare roc curves. BMC bioinformatics 12 (1), 77.

Shi, Q., Abdel-Aty, M., 2015. Big data applications in real-time traffic operation and safety monitoring and improvement on urban expressways. Transportation Research Part C: Emerging Technologies 58, 380-394.

Theofilatos, A., 2017. Incorporating real-time traffic and weather data to explore road accident likelihood and severity in urban arterials. Journal of Safety Research 61, 9-21.

Theofilatos, A., Yannis, G., Kopelias, P., Papadimitriou, F., 2018. Impact of real-time traffic characteristics on crash occurrence: Preliminary results of the case of rare events. Accident Analysis & Prevention.

Theofilatos, A., Yannis, G., Vlahogianni, E.I., Golias, J.C., 2017. Modeling the effect of traffic regimes on safety of urban arterials: The case study of athens. Journal of Traffic and Transportation Engineering (English Edition) 4 (3), 240-251.

Wang, X., Abdel-Aty, M., Almonte, A., Darwiche, A., 2009. Incorporating traffic operation measures in safety analysis at signalized intersections. Transportation Research Record: Journal of the Transportation Research Board (2103), 98-107.

Wang, X., Abdel-Aty, M., Brady, P., 2006. Crash estimation at signalized intersections: Significant factors and temporal effect. Transportation Research Record: Journal of the Transportation Research Board (1953), 10-20.

Wang, X., Yuan, J., 2017. Safety impacts study of roadway network features on suburban highways. China J. Highw. Transp 30, 106-114.

Wang, X., Yuan, J., Schultz, G.G., Fang, S., 2018. Investigating the safety impact of roadway network features of suburban arterials in shanghai. Accident Analysis & Prevention 113, 137-148.

Wang, X., Yuan, J., Yang, X., 2016. Modeling of crash types at signalized intersections based on random effect model. Journal of Tongji University (Natural Science) 44 (1), 81-86.

Xie, K., Wang, X., Huang, H., Chen, X., 2013. Corridor-level signalized intersection safety analysis in shanghai, china using bayesian hierarchical models. Accident Analysis & Prevention 50, 25-33.

Xu, C., Liu, P., Wang, W., Li, Z., 2012. Evaluation of the impacts of traffic states on crash risks on freeways. Accident Analysis & Prevention 47, 162-71.

Xu, C., Tarko, A.P., Wang, W., Liu, P., 2013a. Predicting crash likelihood and severity on freeways with real-time loop detector data. Accident Analysis & Prevention 57, 30-9.

Xu, C., Wang, W., Liu, P., 2013b. Identifying crash-prone traffic conditions under different weather on freeways. Journal of Safety Research 46, 135-44.

Yu, R., Abdel-Aty, M., 2014. Analyzing crash injury severity for a mountainous freeway incorporating real-time traffic and weather data. Safety Science 63, 50-56.

Yu, R., Abdel-Aty, M.A., Ahmed, M.M., Wang, X., 2014. Utilizing microscopic traffic and weather data to analyze real-time crash patterns in the context of active traffic management. IEEE Transactions On Intelligent Transportation Systems 15 (1), 205-213.

Yu, R., Wang, X., Yang, K., Abdel-Aty, M., 2016. Crash risk analysis for shanghai urban expressways: A bayesian semi-parametric modeling approach. Accident Analysis & Prevention 95 (Pt B), 495-502.

Yuan, J., Abdel-Aty, M., Wang, L., Lee, J., Wang, X., Yu, R., Year. Real-time crash risk analysis of urban arterials incorporating bluetooth, weather, and adaptive signal control data. In: Proceedings of the Transportation Research Board 97th Annual MeetingTransportation Research Board, Washington D.C.

Yue, L., Abdel-Aty, M., Wu, Y., Wang, L., 2018. Assessment of the safety benefits of vehicles' advanced driver assistance, connectivity and low level automation systems. Accident Analysis & Prevention 117, 55-64.

Zheng, Z., Ahn, S., Monsere, C.M., 2010. Impact of traffic oscillations on freeway crash occurrences. Accident Analysis & Prevention 42 (2), 626-36.
31